\documentclass[12pt]{article}
\usepackage{amsfonts}
\font\twozero=cmr10 at 20pt
\font\oneeight=cmr10 at 18pt

\font\larl=cmr10 at 24pt
\newcommand{\vT}{\vphantom{\mbox{\twozero I}}}
\newcommand{\vTm}{\vphantom{\mbox{\oneeight I}}}

\newcommand{\vTb}{\vphantom{\mbox{\larl I}}}

\begin{document}

\baselineskip=20pt

%%%%%%%%%%%%%%%%%%%%%%%%%%%%%%%%%%%%%%%%%%%%%%%%%%%%%%%%%%%%
%                                                          %
%  Title page                                              %
%                                                          %
%%%%%%%%%%%%%%%%%%%%%%%%%%%%%%%%%%%%%%%%%%%%%%%%%%%%%%%%%%%%
\newfont{\elevenmib}{cmmib10 scaled\magstep1}
\newcommand{\preprint}{
   \begin{flushleft}
     \elevenmib Yukawa\, Institute\, Kyoto\\
   \end{flushleft}\vspace{-1.3cm}
   \begin{flushright}\normalsize  \sf
     DPSU-02-1\\
     YITP-02-37\\
     {\tt hep-th/0206172} \\ June 2002
   \end{flushright}}
\newcommand{\Title}[1]{{\baselineskip=26pt
   \begin{center} \Large \bf #1 \\ \ \\ \end{center}}}
\newcommand{\Author}{\begin{center}
   \large \bf S.~Odake${}^a$ and R.~Sasaki${}^b$ \end{center}}
\newcommand{\Address}{\begin{center}
     $^a$ Department of Physics, Shinshu University,\\
     Matsumoto 390-8621, Japan\\
     ${}^b$ Yukawa Institute for Theoretical Physics,\\
     Kyoto University, Kyoto 606-8502, Japan
   \end{center}}
\newcommand{\Accepted}[1]{\begin{center}
   {\large \sf #1}\\ \vspace{1mm}{\small \sf Accepted for Publication}
   \end{center}}

\preprint
\thispagestyle{empty}
\bigskip\bigskip\bigskip

\Title{Polynomials Associated with Equilibrium Positions\\
   in Calogero-Moser Systems}
\Author

\Address
\vspace{1cm}

\begin{abstract}
In a previous paper (Corrigan-Sasaki), many remarkable properties of
classical Calogero and Sutherland systems at equilibrium are reported.
For example, the minimum energies, frequencies of small oscillations and
the eigenvalues of Lax pair matrices at equilibrium are all ``integer valued".
The equilibrium positions of Calogero and Sutherland systems for the
classical root systems ($A_r$, $B_r$, $C_r$ and $D_r$) correspond to
the zeros of Hermite, Laguerre, Jacobi and Chebyshev polynomials.
Here we define and derive the corresponding polynomials for the exceptional
($E_6$, $E_7$, $E_8$, $F_4$ and $G_2$) and non-crystallographic ($I_2(m)$,
$H_3$ and $H_4$) root systems. They do not have orthogonality but share
many other properties with the above mentioned classical polynomials.
\end{abstract}

\newpage
%%%%%%%%%%%%%%%%%%%%%%%%%%%%%%%%%%%%%%%%%%%%%%%%%%%%%%%%%%%%%%%
%                                                             %
%  1. Introduction                                            %
%                                                             %
%%%%%%%%%%%%%%%%%%%%%%%%%%%%%%%%%%%%%%%%%%%%%%%%%%%%%%%%%%%%%%%
\section{Introduction}
\label{intro}
\setcounter{equation}{0}

The relationship between classical and quantum integrability
has fascinated many physicists and mathematician.
In a recent paper by Corrigan and Sasaki \cite{cs}, this issue has been
extensively investigated in the framework of Calogero-Moser systems
\cite{Cal,Sut,CalMo}. One major result is that certain
``quantised" information seems to be encoded in the classical
system. For example, the eigenvalues of classical Lax pair matrices
at the equilibrium points are ``integer valued" \cite{cs}.
The connection between the zeros of Hermite and Laguerre
polynomials and the equilibrium points of $A_r$ and $B_r$
($D_r$) Calogero systems has been known for many years
\cite{calmat,calpere,ahmcal}.
In \cite{cs} it is found out that the zeros of Jacobi
polynomials are related to the equilibrium points of $BC_r$
($D_r$) Sutherland system.
In the present paper we define and derive the polynomials associated
with the equilibrium points of the other Calogero and Sutherland 
systems. These are associated with Calogero systems based on
non-crystallographic root systems, Calogero and Sutherland
systems based on the exceptional root systems and the $A_r$ Sutherland
systems. The Chebyshev polynomials (\ref{archev}) are associated with
the $A_r$ Sutherland systems.

In general, the polynomials are determined by the potential,
$q^2+1/q^2$ (the Calogero system \cite{Cal}) and $1/\sin^2q$ (the
Sutherland system \cite{Sut}), the root system $\Delta$
and the set of weights ${\cal R}$. For the classical root systems and for
the (non-trivial) smallest dimensional ${\cal R}$, that is the set of
vector weights ${\bf V}$ or the set of short roots $\Delta_S$,
the polynomials turn out to be classical {\em orthogonal\/}
polynomials; Hermite, Laguerre, Jacobi and Chebyshev polynomials \cite{szego}.
The orthogonality does not hold for the polynomials for the exceptional root
systems and for the classical root systems with generic ${\cal R}$'s.
Like their classical counterparts, these new polynomials
have ``integer coefficients" only, if multiplied by a certain factor.
In most cases, it is possible to define the polynomials to be {\em
monic\/} (that is, the  highest degree term has {\em unit\/}
coefficient) and {\em integer coefficients only}.
Some polynomials are too lengthy to be displayed in the paper;
an $E_8$ polynomial has 121 terms and its  typical integer
coefficient has about 150 digits. They are presented in \cite{poly}.
Some root systems are related by Dynkin diagram foldings;
$A_{2r-1}\to C_r$, $D_{r+1}\to B_r$, $E_6\to F_4$ and $D_4\to G_2$.
These imply relations among the corresponding Calogero-Moser systems
at certain ratio of the coupling constants. These, in turn, imply
relations among the corresponding polynomials, which are determined
independently. These relations are either  identities among
classical polynomials, many of which are ``new" in the sense they are not
listed in standard mathematical textbooks \cite{szego}, or they provide 
non-trivial checks for the newly
derived polynomials. The significance and other detailed properties of
these new polynomials deserve further study.

This paper is organised as follows.
In section two a brief introduction of Calogero-Moser systems is given to
set the stage and notation. Equations for determining equilibrium points
are discussed in some detail.
In section three Coxeter (Weyl) invariant polynomials associated with
equilibrium positions are introduced for a set of weights ${\cal R}$
for Calogero and Sutherland systems. For the rational potential (Calogero
systems) the definition is almost unique, whereas we have several choices
of the definitions of the polynomials for the trigonometric potential
(Sutherland system). Section four and five are the main body of the paper.
The Coxeter (Weyl) invariant polynomials are determined and presented for all
root systems $\Delta$ and for major choices of ${\cal R}$'s for Calogero
(section four) and Sutherland systems (section five). Section six
is for summary and comments. We will present a heuristic argument for
deriving the classical orthogonal polynomials starting from the
pre-potentials (\ref{Wform}) of Calogero and Sutherland systems.

%%%%%%%%%%%%%%%%%%%%%%%%%%%%%%%%%%%%%%%%%%%%%%%%%%%%%%%%%%%%%%%
%                                                             %
%  2. Equilibrium in Calogero-Moser System                    %
%                                                             %
%%%%%%%%%%%%%%%%%%%%%%%%%%%%%%%%%%%%%%%%%%%%%%%%%%%%%%%%%%%%%%%
\section{Equilibrium in Calogero-Moser System}
\label{calmo}
\setcounter{equation}{0}

Let us start with a brief introduction of
Calogero-Moser systems \cite{Cal,Sut,CalMo}.
We stick to the notation of a recent paper \cite{cs},
unless otherwise mentioned.
Calogero-Moser systems are integrable multiparticle dynamical systems
at the classical as well as  quantum levels.
They have a long range potential (rational, trigonometric,
hyperbolic and elliptic) and the integrable multiparticle interactions
are governed by the root systems \cite{OP1}.
Classical integrability through Lax formalism is known for all potentials
for the classical root systems \cite{OP1} as well as for the exceptional
\cite{DHoker_Phong,bcs2} and non-crystallographic \cite{bcs2} root systems.
Quantum integrability of the systems having degenerate potentials
(rational, trigonometric and hyperbolic) is now systematically understood
for all root systems in terms of Dunkl operator formalism \cite{Dunk,HeOp}
and the quantum Lax pair formalism \cite{bms,kps}.
To a system of $r$ particles in one dimension, we associate a root system
$\Delta$  of rank \(r\). This is a set of vectors in \(\mathbb{R}^{r}\)
invariant under reflections in the hyperplane perpendicular to each
vector in $\Delta$:
\begin{equation}
   \Delta\ni s_{\alpha}(\beta)=\beta-(\alpha^{\vee}\cdot\beta)\alpha,
   \quad \alpha^{\vee}={2\alpha\over{\alpha^2}},
   \quad \alpha, \beta\in\Delta.
   \label{a1}
\end{equation}
The set of reflections $\{s_{\alpha}|\,\alpha\in\Delta\}$ generates a
finite reflection group $G_{\Delta}$, known as a Coxeter (or Weyl) group.
Among Calogero-Moser systems the Calogero systems (with $q^2+1/q^2$
potential) and the Sutherland systems (with $1/\sin^2\!q$ potential) have
discrete energy eigenvalues only when quantised.
The Calogero and Sutherland systems have equilibrium positions,
which are characterised by two equivalent ways \cite{cs}. That is where
the classical potential takes the absolute minimum and simultaneously
the groundstate wavefunction takes the absolute maximum.
At the equilibrium positions of the Calogero and Sutherland systems,
associated spin exchange models are defined for each root system \cite{is1}.
The best known example is  the Haldane-Shastry model which is based on
$A_r$ Sutherland systems \cite{halsha}.
The integrability and the well ordered spectrum of the spin exchange
models are closely related with the special properties of
systems at equilibrium \cite{cs}.

The classical Hamiltonians of the Calogero and Sutherland systems read%
\footnote{For $\Delta=BC_r$ the trigonometric potential should read
$g_{M}^2\sum_{\rho\in\Delta_{M+}}1/\sin^2(\rho\cdot q)
+2g_{L}^2 \sum_{\rho\in\Delta_{L+}}1/\sin^2(\rho\cdot q)
+g_{S}(g_S+2g_L)/2\,\sum_{\rho\in\Delta_{S+}}1/ \sin^2(\rho\cdot q)$,
with $\rho_M^2=2$, $\rho_L^2=4$ and $\rho_S^2=1$.}:
\begin{eqnarray}
   &&{\cal H}_C={1\over 2} p^{2} +V_C,\quad
   V_C=\left\{
   \begin{array}{l}
   \begin{displaystyle}
     {\omega^2\over2}q^2 + {1\over2}\sum_{\rho\in\Delta_+}
     {g_{\rho}^2 \rho^{2}\over{(\rho\cdot q)^2}}\,,
   \end{displaystyle}\\[20pt]
   \begin{displaystyle}
   {1\over2}\sum_{\rho\in\Delta_+}
   {g_{\rho}^2 \rho^{2}\over{\sin^2(\rho\cdot q)}}\,.
   \end{displaystyle}
   \end{array}\right.
   \label{CHam}
\end{eqnarray}
In these formulae, $\Delta_+$ is the set of positive roots and
$\omega>0$ is the angular frequency of the confining harmonic potential,
\(g_{\rho}>0\) are real coupling constants which are defined on orbits of
the corresponding Coxeter group, {\it i.e.,} they are identical for roots
in the same orbit.
The classical potential $V_C$ can be written succinctly
in terms of a {\em pre-potential\/} $W$ \cite{bms}:
\begin{eqnarray}
   &&V_C={1\over2}\sum_{j=1}^r\left(
   {\partial W\over{\partial q_j}}\right)^2+\tilde{\cal E}_0,
   \label{cpot}
\end{eqnarray}
in which
\begin{eqnarray}
   &&W=\left\{
   \begin{array}{c}
   \begin{displaystyle}
   -{\omega\over2}q^2+\sum_{\rho\in\Delta_+}g_{\rho}\log|\rho\cdot q|\,,
   \end{displaystyle}\\[16pt]
   \begin{displaystyle}
   \sum_{\rho\in\Delta_+}g_{\rho}\log|\sin(\rho\cdot q)|\,,
   \end{displaystyle}
   \end{array}\right.
   \label{Wform}
\end{eqnarray}
and $\tilde{\cal E}_0$ is the minimum energy.
Let us recall that the pre-potential $W$ is related to the
ground state wavefunction of the quantum  theory $\phi_0$ by $\phi_0=e^W$
(eq.(2.6) of \cite{cs})
and that $W$, $V_C$ and ${\cal H}_C$ are Coxeter (Weyl) invariant:
\begin{equation}
   {\cal H}_C(p,q)={\cal H}_C(s_\alpha(p),s_\alpha(q)),\;\;
   W(q)=W(s_\alpha(q)),\;\; V_C(q)=V_C(s_\alpha(q))\;\;\;
   (\forall\alpha\in\Delta).
   \label{wCoxinv}
\end{equation}

The classical equilibrium point
\begin{equation}
   p=0,\quad q=\bar{q}
   \label{clasequil}
\end{equation}
is determined by the equations \cite{cs}
\begin{equation}
   \left.{\partial V_C\over{\partial q_j}}\right|_{\bar{q}}=0
   \mbox{~~or equivalently~~}
   \left.{\partial W\over{\partial q_j}}\right|_{\bar{q}}=0\quad(j=1,\ldots,r).
   \label{Wmax}
\end{equation}
In other words, it is a {\em minimal\/} point of the classical
potential $V_C$, and simultaneously it is a {\em maximal\/} point of
the pre-potential $W$ and of the ground state  wavefunction $\phi_0=e^W$,
since the matrix determining the frequencies of small oscillations around
the equilibrium
\begin{equation}
   \left.{\partial^2 W\over{\partial q_j \partial q_l}}\right|_{\bar{q}}
   \quad (j,l=1,\ldots, r),
   \label{wpp}
\end{equation}
is negative definite \cite{cs}.
The equilibrium points are not unique.
There is one equilibrium point in each Weyl chamber (alcove) \cite{cs},
that is if $\bar{q}$ is an equilibrium point, so is $s_\rho(\bar{q})$,
$\forall\rho\in\Delta$, due to the Coxeter (Weyl) invariance of $W$
(\ref{wCoxinv}). It is also easy to see that if $\bar{q}$ is an
equilibrium point, so is $-\bar{q}$.

The equilibrium equation for the pre-potential $W$,
for Calogero systems based on {\em simply laced\/} root systems,
that is $A_r$, $D_r$, $E_r$, $I_2(\rm odd)$ and $H_r$, reads:
\[
   \sum_{\rho\in\Delta_+}{\rho\over{\rho\cdot\bar{q}}}={\omega\over{g}}\bar{q}.
\]
If we define a {\em rescaled equilibrium point\/} by
\begin{equation}
   \tilde{q}\equiv \sqrt{\omega\over g}\,\bar{q},
\end{equation}
it satisfies a simple equation independent of the coupling constant:
\begin{equation}
   \sum_{\rho\in\Delta_+}{\rho\over{\rho\cdot\tilde{q}}}=\tilde{q}.
   \label{rsleq}
\end{equation}
For Calogero systems based on {\em non-simply laced\/} root systems,
that is $B_r$, $C_r$, $F_4$, $G_2$ and $I_2(\rm even)$%
\footnote{For $I_2(\rm even)$ we have $k\equiv g_e/g_o$.},
the equation reads:
\[
   \sum_{\rho\in\Delta_{L+}}{\rho\over{\rho\cdot\bar{q}}}+
   k\sum_{\rho\in\Delta_{S+}}{\rho\over{\rho\cdot\bar{q}}}=
   {\omega\over{g_L}}\bar{q},
   \qquad k\equiv{g_S\over{g_L}}.
\]
Again a {\em rescaled equilibrium point\/}
\begin{equation}
   \tilde{q}\equiv \sqrt{\omega\over g_L}\,\bar{q},
\end{equation}
satisfies a simple equation depending only on the ratio
of the two coupling constants $g_S$ and $g_L$:
\begin{equation}
   \sum_{\rho\in\Delta_{L+}}{\rho\over{\rho\cdot\tilde{q}}}+
   k\sum_{\rho\in\Delta_{S+}}{\rho\over{\rho\cdot\tilde{q}}}=\tilde{q},
   \qquad k\equiv{g_S\over{g_L}}.
   \label{rnsleq}
\end{equation}
As is clear from (\ref{rsleq}) and (\ref{rnsleq}), the equilibrium
point $\tilde{q}$ ($\bar{q}$) is independent of the normalisation of
roots in $\Delta$.

The situation is simpler in the Sutherland systems which do not have an
extra parameter $\omega$. For crystallographic {\em simply laced\/}
root systems, that is $A_r$, $D_r$ and $E_r$, the equation for $\bar{q}$ is
independent of the coupling constant:
\begin{equation}
   \sum_{\rho\in\Delta_+}\!{\rho\cot(\rho\cdot\bar{q})}=0.
   \label{tsleq}
\end{equation}
For crystallographic {\em non-simply laced\/} root systems, that is
$B_r$, $C_r$, $F_4$ and $G_2$, the equation for $\bar{q}$ depends only
on the ratio of the two coupling constants $g_S$ and $g_L$:
\begin{equation}
   \sum_{\rho\in\Delta_{L+}}\!{\rho\cot(\rho\cdot\bar{q})}+
   k\sum_{\rho\in\Delta_{S+}}\!{\rho\cot(\rho\cdot\bar{q})}=0,
   \qquad k\equiv{g_S\over{g_L}}.
   \label{tnsleq}
\end{equation}
For the $BC_r$ system, which has {\em three\/} coupling constants
$g_S$, $g_M$ and $g_L$ for the short, middle and long roots,
the equation depends on two coupling ratios:
\begin{equation}
   \sum_{\rho\in\Delta_{M+}}\!{\rho\cot(\rho\cdot\bar{q})}+
   k_1\sum_{\rho\in\Delta_{S+}}\!{\rho\cot(\rho\cdot\bar{q})}+
   k_2\sum_{\rho\in\Delta_{L+}}\!{\rho\cot(\rho\cdot\bar{q})}=0,
   \quad k_1\equiv{g_S\over{g_M}}, \quad k_2\equiv{g_L\over{g_M}}.
   \label{tbcreq}
\end{equation}

%%%%%%%%%%%%%%%%%%%%%%%%%%%%%%%%%%%%%%%%%%%%%%%%%%%%%%%%%%%%%%%
%                                                             %
%  3. Polynomials                                             %
%                                                             %
%%%%%%%%%%%%%%%%%%%%%%%%%%%%%%%%%%%%%%%%%%%%%%%%%%%%%%%%%%%%%%%
\section{Polynomials}
\label{polydef}
\setcounter{equation}{0}

Here we give the general definitions of the Coxeter (Weyl) invariant
polynomials associated with equilibrium positions in Calogero and
Sutherland systems. Naturally, the definitions for the Calogero systems
are different from those for the Sutherland systems except for the common
features that the polynomials are Coxeter (Weyl) invariant and are
specified by the root system $\Delta$ and a set of $D$ vectors ${\cal R}$
\begin{equation}
   {\cal R}=\{\mu^{(1)},\ldots,\mu^{(D)}\,|\,\mu^{(a)}\in {\mathbb R}^r\},
\end{equation}
which form a single orbit of the corresponding reflection (Weyl) group
$G_\Delta$. The set of values at the equilibrium,
$\{\mu\cdot\bar{q}\,|\,\mu\in{\cal R}\}$, is Coxeter (Weyl) invariant.
In this paper we consider only such ${\cal R}$'s that are customarily
used for Lax pairs.
They are the set of roots $\Delta$ itself for simply laced root systems,
the set of long (short, middle) roots $\Delta_L$ ($\Delta_S$, $\Delta_M$)
for non-simply laced root  systems and the so-called sets of
{\em minimal weights\/}.
The latter is better specified by the corresponding fundamental
representations, which are all the fundamental representations of $A_r$,
the vector ($\bf V$), spinor ($\bf S$) and conjugate spinor ($\bar{\bf S}$)
representations of $D_r$ and $\bf 27$ ($\overline{\bf 27}$) of $E_6$
and $\bf 56$ of $E_7$.

For Calogero systems the definition is rather unique and  it is given by
\begin{equation}
   P_{\Delta}^{\cal R}(k |x)=\prod_{\mu\in{\cal R}}(x-\mu\cdot\tilde{q}),
   \label{calpolydef}
\end{equation}
in which $k$ denotes the possible dependence on the ratio of the coupling
constants, for the systems based non-simply laced root systems (\ref{rnsleq}).
It should be noted that the above polynomial depends on the normalisation
of the vectors $\mu\in{\cal R}$ implicitly. Changing ${\cal R}\to c{\cal R}$
($\mu\to c\mu$) can be absorbed by rescaling of $x$:
\begin{equation}
   P_{\Delta}^{c\cal R}(k |x)=\prod_{\mu\in{\cal R}}(x-c\mu\cdot\tilde{q})=
   c^DP_{\Delta}^{\cal R}(k |x/c).
\end{equation}

For Sutherland systems we have several candidates for polynomials:
\begin{eqnarray}
   &&P_{\Delta,\, s}^{\cal R}(k|x)=
   \prod_{\mu\in{\cal R}}\Bigl(x-\sin(\mu\cdot\bar{q})\Bigr),\quad
   P_{\Delta,\, s2}^{\cal R}(k|x)=
   \prod_{\mu\in{\cal R}}\Bigl(x-\sin(2\mu\cdot\bar{q})\Bigr),
   \label{sutpolysdef}\\
   &&P_{\Delta,\, c}^{\cal R}(k|x)=
   \prod_{\mu\in{\cal R}}\Bigl(x-\cos(\mu\cdot\bar{q})\Bigr),\quad
   P_{\Delta,\, c2}^{\cal R}(k|x)=
   \prod_{\mu\in{\cal R}}\Bigl(x-\cos(2\mu\cdot\bar{q})\Bigr),
   \label{sutpolycdef}
\end{eqnarray}
in which $k$ denotes possible dependence on the ratio(s) of coupling constants,
as before. Not all of them give interesting objects, as we will see presently.
In all cases the polynomials are monic and of degree $D$.

In case ${\cal R}$ is {\em even\/}, that is,
\begin{equation}
   \mu\in{\cal R}\Longleftrightarrow -\mu\in{\cal R},
\end{equation}
then sometimes it is advantageous to consider
$P_{\Delta}^{\cal R}(k|x)$, $P_{\Delta,\, s}^{\cal R}(k|x)$ and
$P_{\Delta,\, s2}^{\cal R}(k|x)$ as polynomials in $y\equiv x^2$ of
degree $D/2$:
\begin{equation}
   \prod_{\mu\in{\cal R_+}}\left(y-(\mu\cdot\tilde{q})^2\right), \quad
   \prod_{\mu\in{\cal R_+}}\Bigl(y-\sin^2(\mu\cdot\bar{q})\Bigr),
   \quad \prod_{\mu\in{\cal R_+}}\Bigl(y-\sin^2(2\mu\cdot\bar{q})\Bigr),
   \label{evenpol}
\end{equation}
in which ${\cal R}_+$ is the {\em positive\/} part of ${\cal R}$.
In this case the ``cosine" polynomials $P_{\Delta,\, c(2)}^{\cal R}(k|x)$,
(\ref{sutpolycdef}) should better be redefined as
\begin{eqnarray}
   &&P_{\Delta,\, c}^{\cal R_+}(k|x)=
   \prod_{\mu\in{\cal R}_+}\Bigl(x-\cos(\mu\cdot\bar{q})\Bigr),\quad
   P_{\Delta,\, c2}^{\cal R_+}(k|x)=
   \prod_{\mu\in{\cal R}_+}\Bigl(x-\cos(2\mu\cdot\bar{q})\Bigr),
   \label{sutpolycdefplus}
\end{eqnarray}
since the original polynomials (\ref{sutpolycdef}) are the squares of
the new ones. It is easy to see that $P_{\Delta,\, s}^{\cal R}(k|y)$ and
$P_{\Delta,\, c2}^{\cal R_+}(k|x)$ are equivalent:
\begin{equation}
   P_{\Delta,\, s}^{\cal R}(k|x)=(-2)^{-D/2}
   P_{\Delta,\, c2}^{\cal R_+}(k|1-2x^2).
   \label{sc2rel}
\end{equation}
Likewise, for {\em even\/} ${\cal R}$, $P_{\Delta,\, s2}^{\cal R}(k|x)$
is a ``square" of  $P_{\Delta,\, c2}^{{\cal R}_+}(k|x)$:
\begin{eqnarray}
   P_{\Delta,\, s2}^{\cal R}(k|x)&\!\!=\!\!&
   \prod_{\mu\in{\cal R}}\Bigl(x-\sin(2\mu\cdot\bar{q})\Bigr)
   =\prod_{\mu\in{{\cal R}_+}}\Bigl(x^2-\sin^2(2\mu\cdot\bar{q})\Bigr)
   \nonumber\\
   &\!\!=\!\!&
   \prod_{\mu\in{{\cal R}_+}}\Bigl(u-\cos(2\mu\cdot\bar{q})\Bigr)
   \Bigl(-u-\cos(2\mu\cdot\bar{q})\Bigr),
   \quad u^2\equiv 1-x^2\nonumber\\
   &\!\!=\!\!&
   P_{\Delta,\, c2}^{{\cal R}_+}(k|u)P_{\Delta,\, c2}^{{\cal R}_+}(k|-u).
   \label{s2c2rel}
\end{eqnarray}
The right hand side is an even polynomial in $u$, thus it is a polynomial
in $u^2$ and in $x^2$.
The change of variables $u\leftrightarrow x$ corresponds to the change
in the character of the variables, $\cos \leftrightarrow \sin$.
This imposes a quite non-trivial check for the $s2$
and $c2$ polynomials which are determined separately.

As shown in the following sections, the polynomials associated with the
classical root systems ($A_r$, $B_r$, $C_r$ and $D_r$) and $I_2(m)$
are either classical polynomials for the smallest dimensional ${\cal R}$ or
those closely related to them, see for example, (\ref{rfgrel1}),
(\ref{rfgrel2}), (\ref{tfgrel1}), (\ref{tfgrel2}).
For the exceptional and non-crystallographic root systems, the equilibrium
positions are evaluated numerically and the polynomials are obtained by
rationalisation of the coefficients in terms of Mathematica.
At each step, the result is verified by many consistency checks;
the ``integer eigenvalues" of the matrix (\ref{wpp}) for the values of
$\bar{q}$, the identities implied by Dynkin diagram foldings and
identities (\ref{s2c2rel}) for the polynomials.
Let us conclude this section with an important remark that these polynomials
are independent of the specific representation of the root and weight vectors.
In other words, the polynomials are Coxeter (Weyl) invariant.

%%%%%%%%%%%%%%%%%%%%%%%%%%%%%%%%%%%%%%%%%%%%%%%%%%%%%%%%%%%%%%%
%                                                             %
%  4. Calogero Systems                                        %
%                                                             %
%%%%%%%%%%%%%%%%%%%%%%%%%%%%%%%%%%%%%%%%%%%%%%%%%%%%%%%%%%%%%%%
\section{Calogero Systems}
\label{calsys}
\setcounter{equation}{0}

Let us first discuss the systems based on the classical root systems.
%%%%%%%%%%%%%%%%%%%%%%%%%%%%
%  4.1                     %
%%%%%%%%%%%%%%%%%%%%%%%%%%%%
\subsection{$A_r$}

The equations (\ref{rsleq}) for $\Delta=A_r$ read
\begin{equation}
   \sum_{l=1\atop l\ne j}^{r+1}{1\over{\tilde{q}_j-\tilde{q}_l}}=\tilde{q}_j,
   \quad (j=1,\ldots,r+1).
   \label{herroots}
\end{equation}
These determine $\{\tilde{q}_j=\sqrt{\omega\over g}\bar{q}_j\,|\,
j=1,\ldots,r+1\}$ to be the zeros of the Hermite polynomial $H_{r+1}(x)$
\cite{szego}, with the Rodrigues' formula
\begin{equation}
   H_n(x)=(-1)^ne^{x^2}\left({d\over{dx}}\right)^ne^{-x^2}
   =2^nx^n+\cdots.
\end{equation}

If ordered by the value, $\tilde{q}_1>\tilde{q}_2>\cdots>\tilde{q}_{r+1}$
or reverse, they possess the symmetry
\begin{equation}
   \tilde{q}_j=-\tilde{q}_{r+2-j},
   \label{qj-qr+2-j}
\end{equation}
and especially $\tilde{q}_{(r+2)/2}=0$ for $r$ even. Thus we have
\begin{equation}
   \tilde{q}_1+\tilde{q}_2+\cdots+\tilde{q}_{r+1}=0.
   \label{sumzero}
\end{equation}

%%%%%%%%%%%%%%%%%%%%%%%%%%%%
%  4.1.1                   %
%%%%%%%%%%%%%%%%%%%%%%%%%%%%
\subsubsection{${\cal R}={\bf V}$ for $A_r$}

This case was reported by Calogero a quarter century ago \cite{calmat}.
The set of weights of the vector representation is
\begin{equation}
   {\bf V}=\Bigl\{\mu_j\equiv{\bf e}_j-\frac{1}{r+1}\sum_{l=1}^{r+1}{\bf e}_l
   \Bigm|j=1,\ldots,r+1\Bigr\}.
   \label{ArV}
\end{equation}
Throughout this paper we denote an orthonormal basis of ${\mathbb R}^r$
(${\mathbb R}^{r+1}$ for
$A_r$ case) by $\{{\bf e}_j\}$.

In this case, we have $\mu_j\cdot\tilde{q}=\tilde{q}_j$ due to (\ref{sumzero})
and $\mu^2=r/(r+1)$.
The polynomial (\ref{calpolydef}) is given by the Hermite polynomial
\begin{equation}
   P_r^{\bf V}(x)\equiv P_{A_r}^{\bf V}(x)=
   \prod_{j=1}^{r+1}\left(x-\tilde{q}_j\right)
   =2^{-(r+1)}H_{r+1}(x).
   \label{hermpoly}
\end{equation}
They are orthogonal to each other:
\begin{equation}
   \int_{-\infty}^\infty P_r^{\bf V}(x)P_s^{\bf V}(x)\,e^{-x^2}\,dx\,
   \propto\delta_{r\,s}.
\end{equation}
Needless to say, Hermite polynomials are of integer coefficients.
It is interesting to note that another definition
\begin{equation}
   P_{A_r}^{2{\bf V}}(x)=
   \prod_{j=1}^{r+1}\left(x-2\tilde{q}_j\right)
   =H_{r+1}(x/2)=2^{r+1}P_{A_r}^{\bf V}(x/2).
   \label{hermpoly2}
\end{equation}
gives a {\em monic} polynomial with {\em all integer coefficients\/}.

%%%%%%%%%%%%%%%%%%%%%%%%%%%%
%  4.1.2                   %
%%%%%%%%%%%%%%%%%%%%%%%%%%%%
\subsubsection{${\cal R}={\bf V}_i$ for $A_r$}

The set of weights of the $i$-th fundamental representation (\,$i$-th rank
anti-symmetric tensor representation, $1\leq i\leq r$) is
\begin{equation}
   {\bf V}_i=\Bigl\{\mu_{j_1}+\cdots+\mu_{j_i}\Bigm|
   1\leq j_1<\cdots<j_i\leq r+1\Bigr\},\quad D=D_i\equiv{r+1\choose i}.
   \label{ArViset}
\end{equation}
The above ${\bf V}$ (\ref{ArV}) is ${\bf V}={\bf V}_1$.
In this case we have $\mu^2=i(r+1-i)/(r+1)$.
We can show that the polynomial (\ref{calpolydef})
\begin{equation}
   P_{A_r}^{{\bf V}_i}(x)=\prod_{1\leq j_1<\cdots<j_i\leq r+1}
   \Bigl(x-(\tilde{q}_{j_1}+\cdots\tilde{q}_{j_i})\Bigr)
   =P_{A_r}^{{\bf V}_{r+1-i}}(x)
   \label{ArVi}
\end{equation}
can be expressed in terms of the coefficients of $H_{r+1}(x)$
by the same method as given in section \ref{brrdell}, and
$P_{A_r}^{2{\bf V}_i}(x)$ gives a monic polynomial with integer coefficients.

Here we report only on ${\bf V}_2$ because it seems that the other 
representations ($3\leq i\leq r-2$) do not provide any interesting results.
(For lower rank $r$, the explicit forms of the polynomials 
$P_{A_r}^{{\bf V}_i}(x)$ can be found in \cite{poly}.)
Due to (\ref{qj-qr+2-j}), eq.(\ref{ArVi}) becomes
\begin{eqnarray}
   P_{A_r}^{{\bf V}_2}(x)&\!\!=\!\!&
   \prod_{1\leq j<\,l\,\leq r+1}
   \Bigl(x-(\tilde{q}_j+\tilde{q}_l)\Bigr)\\
   &\!\!=\!\!&
   \left\{\begin{array}{ll}
   {\displaystyle x^{(r+1)/2}\prod_{1\leq j<\,l\,\leq(r+1)/2}
   \Bigl(x^2-(\tilde{q}_j-\tilde{q}_l)^2\Bigr)
   \Bigl(x^2-(\tilde{q}_j+\tilde{q}_l)^2\Bigr)}&\mbox{$r$ : odd}\\
   {\displaystyle x^{r/2}\prod_{j=1}^{r/2}(x^2-\tilde{q}_j^2)
   \;\cdot\!\!\!\!\! \prod_{1\leq j<\,l\,\leq r/2}
   \Bigl(x^2-(\tilde{q}_j-\tilde{q}_l)^2\Bigr)
   \Bigl(x^2-(\tilde{q}_j+\tilde{q}_l)^2\Bigr)}&\mbox{$r$ : even}.
   \end{array}\right.\nonumber
\end{eqnarray}
Based on the fact that the zeros of Hermite and Laguerre
polynomials are related as seen from the formulae (\ref{hermlagrel}),
this can be expressed by using the polynomials associated with the $B_r$
Calogero systems in the following way:
\begin{equation}
   P_{A_{2r-1}}^{{\bf V}_2}(x)=x^rP_{B_r}^{\Delta_L}(1/2|x),\quad
   P_{A_{2r}}^{{\bf V}_2}(x)=
   x^{r-1}P_{A_{2r}}^{{\bf V}}(x)P_{B_r}^{\Delta_L}(3/2|x).
\end{equation}
The explicit forms of the functions $P_{B_r}^{\Delta_L}(k|x)$ for lower
$r$ are given in section \ref{brrdell}.

%%%%%%%%%%%%%%%%%%%%%%%%%%%%
%  4.1.3                   %
%%%%%%%%%%%%%%%%%%%%%%%%%%%%
\subsubsection{${\cal R}=\Delta$ for $A_r$}

We have $\Delta=\{\pm({\bf e}_j-{\bf e}_l)\,|\,1\leq j<l\leq r+1\}$,
$D=r(r+1)$ and $\mu^2=2$.
The polynomial has a factorized form:
\begin{equation}
   P_{A_r}^{\Delta}(x)=
   \prod_{1\leq j<l\leq r+1}\left(\vTm x^2-(\tilde{q}_j-\tilde{q}_l)^2\right)
   =\left\{\begin{array}{ll}
    x^2-2&\;(r=1)\\
    {\displaystyle
    x^{-r-1}P_{A_r}^{2{\bf V}}(x)\Bigl(P_{A_r}^{{\bf V}_2}(x)\Bigr)^2}
    &\;(r\geq 2)\,.
   \end{array}\right.
\end{equation}
Another definition $P_{A_r}^{2\Delta}(x)$ gives a monic polynomial with 
integer coefficients.

%%%%%%%%%%%%%%%%%%%%%%%%%%%%
%  4.2                     %
%%%%%%%%%%%%%%%%%%%%%%%%%%%%
\subsection{$B_r$ and $D_r$}

Assuming $\bar{q}_j\ne0$, the equations (\ref{rnsleq}) for $\Delta=B_r$
with $k\equiv g_S/g_L$ read
\begin{equation}
   \sum_{l=1\atop l\ne j}^{r}{1\over{\tilde{q}_j^2-\tilde{q}_l^2}}
   +{k\over{2\tilde{q}_j^2}}={1\over{2}}\quad(j=1,\ldots,r).
   \label{lagroots}
\end{equation}
They determine $\{\tilde{q}_j^2={\omega\over g_L}\bar{q}_j^2\,|\,
j=1,\ldots,r\}$, as the zeros of the associated Laguerre polynomial
$L_{r}^{(\alpha)}(x)$, with $\alpha=k-1=g_S/g_L-1$ \cite{cs,szego,OP1}.
The Rodrigues' formula reads
\begin{equation}
   L_n^{(\alpha)}(x)={e^x x^{-\alpha}\over{n!}}\left({d\over{dx}}\right)^n
   \left(e^{-x}\,x^{n+\alpha}\right)=\frac{(-1)^n}{n!}x^n+\cdots.
\end{equation}
For the subcase with $g_S=0$, that is $\Delta=D_r$, $\{\tilde{q}_j^2=
{\omega\over g_L}\bar{q}_j^2\,|\,j=1,\ldots,r\}$, are the zeros of the
associated Laguerre  polynomial \cite{szego,OP1},
\begin{equation}
   r\,L_{r}^{(-1)}(x)=-xL_{r-1}^{(1)}(x),
   \label{lagiden}
\end{equation}
for which one of the $\tilde{q}_j$ is zero.
This also means that the $\{\tilde{q}_j^2\}$ of $B_r$ for $g_S/g_L=2$ or
$\alpha=1$ are the same as the non-vanishing $\{\tilde{q}_j^2\}$ of $D_{r+1}$.
This can be understood easily from the Dynkin diagram folding $D_{r+1}\to B_r$.
We omit $C_r$ case, because $C_r$ is obtained from $B_r$ by interchanging the
short ($S$) and long ($L$) roots.

%%%%%%%%%%%%%%%%%%%%%%%%%%%%
%  4.2.1                   %
%%%%%%%%%%%%%%%%%%%%%%%%%%%%
\subsubsection{${\cal R}=\Delta_S$ for $B_r$}

Since $\Delta_S=\{\pm{\bf e}_j\,|\,j=1,\ldots,r\}$ is {\em even\/},
it is advantageous to consider the polynomials in $y\equiv x^2$,
(\ref{evenpol}),
\begin{equation}
   P_r^{\Delta_S}(y)\equiv P_{B_r}^{\Delta_S}(k|x)=
   \prod_{j=1}^{r}\left(x^2-\tilde{q}_j^2\right)
   =(-1)^rr!L_{r}^{(\alpha)}(y),\quad \alpha=k-1=g_S/g_L-1.
   \label{lagpoly}
\end{equation}
They are orthogonal to each other:
\begin{equation}
   \int_{0}^\infty P_r^{\Delta_S}(y)P_s^{\Delta_S}(y)\,y^\alpha e^{- y}\,dy\,
   \propto\delta_{r\,s}.
\end{equation}
It should be stressed that $P_r^{\Delta_S}(y)$, a {\em monic\/} polynomial
in $y$, is also a polynomial in the parameter $\alpha$ with
{\em all integer coefficients\/}.

%%%%%%%%%%%%%%%%%%%%%%%%%%%%
%  4.2.2                   %
%%%%%%%%%%%%%%%%%%%%%%%%%%%%
\subsubsection{${\cal R}={\bf V}$ for $D_r$}

As in the previous example, ${\bf V}=\{\pm{\bf e}_j\,|\,j=1,\ldots,r\}$,
we introduce ($y\equiv x^2$, (\ref{evenpol}))
\begin{equation}
   P_r^{\bf V}(y)\equiv P_{D_r}^{\bf V}(x)=
   \prod_{j=1}^{r}\left(x^2-\tilde{q}_j^2\right)
   =(-1)^rr! L_{r}^{(-1)}(y).
\end{equation}
They are orthogonal to each other:
\begin{equation}
   \int_{0}^\infty P_r^{\bf V}(y)P_s^{\bf V}(y)\,y^{-1} e^{-y}\,dy\,
   \propto \int_{0}^\infty L_{r-1}^{(1)}(y)L_{s-1}^{(1)}(y)
   y\, e^{-y}\,dy\,
   \propto\delta_{r\,s},
\end{equation}
in which the identity (\ref{lagiden}) is used.
Corresponding to the above mentioned Dynkin diagram folding
$D_{r+1}\to B_r$ and (\ref{lagiden}), we obtain
\begin{equation}
   x^2P_{B_r}^{\Delta_S}(2|x)=P_{D_{r+1}}^{\bf V}(x)
   =P_{B_{r+1}}^{\Delta_S}(0|x).
   \label{drbrredV}
\end{equation}

%%%%%%%%%%%%%%%%%%%%%%%%%%%%
%  4.2.3                   %
%%%%%%%%%%%%%%%%%%%%%%%%%%%%
\subsubsection{$A_{2r-1}\to C_r$ and the relationship between Hermite
and Laguerre polynomials}

As is well-known the Dynkin diagram folding $A_{2r-1}\to C_r$ relates
the $A_{2r-1}$ Calogero system to the $C_r$ ($B_r$) system with
$\omega\to2\omega$, $g_S(g_L)=2g$ and $g_L(g_S)=g$, that is $\alpha=-1/2$.
This would imply $P_{A_{2r-1}}^{\bf V}(x)$ (\ref{hermpoly}) is equal to
$P_{B_r}^{\Delta_S}(1/2|x)$ (\ref{lagpoly}):
\begin{equation}
   P_{A_{2r-1}}^{\bf V}(x)=P_{B_r}^{\Delta_S}(1/2|x),
   \label{a2r-1br}
\end{equation}
which is equivalent to a well-known formula relating Hermite polynomials
and Laguerre polynomials (eq(5.6.1) of \cite{szego}):
\begin{equation}
   H_{2r}(x)=(-1)^r2^{2r}r!L_r^{(-1/2)}(x^2),\quad
   H_{2r+1}(x)=(-1)^r2^{2r+1}r!xL_r^{(1/2)}(x^2).
   \label{hermlagrel}
\end{equation}
The former corresponds to $k=1/2$ and (\ref{a2r-1br}). The latter
corresponds to $k=3/2$ and implies
\begin{equation}
   P_{A_{2r}}^{\bf V}(x)=xP_{B_r}^{\Delta_S}(3/2|x).
\end{equation}
Let us recall the corresponding results in the trigonometric case
\cite{szego,cs}.
The polynomial $P_{BC_r,\,c2}^{\Delta_{S+}}(k_1,\,k_2|x)$ (\ref{bcsjac})
is proportional to Jacobi polynomial $P_{r}^{(\alpha,\beta)}(x)$
with $\alpha=k_1+k_2-1$ and $\beta=k_2-1$.
For $k_1=0$, $k_2=1/2$ ($k_1=0$, $k_2=3/2$) it reduces to the Chebyshev
polynomial of the first (second) kind. As above,
$k_1=0$, $k_2=1/2$ corresponds to the $A_{2r-1}\to C_r$ folding.

%%%%%%%%%%%%%%%%%%%%%%%%%%%%
%  4.2.4                   %
%%%%%%%%%%%%%%%%%%%%%%%%%%%%
\subsubsection{${\cal R}={\bf S}$ and $\bar{\bf S}$ for $D_r$}

The spinor ${\bf S}$ and conjugate spinor $\bar{\bf S}$ representations
of $D_r$ are minimal representations with $D=2^{r-1}$ and the natural
normalisation $\mu^2=r/4$.
For odd $r$, we have the equality $-{\bf S}=\bar{\bf S}$ which means
$P_{D_{r}}^{\bf S}(x)=P_{D_{r}}^{\bar{\bf S}}(x)$ for odd $r$.
In fact, the symmetry of the $D_r$ Dynkin diagram implies that
the same formula holds for even $r$, too.
Here we present $P_{D_r}^{\bf S}(x)$ for lower members of $r$:
\begin{eqnarray}
   P_{D_{4}}^{{\bf S},\,\bar{\bf S},\,{\bf V}}(x)&\!\!=\!\!&
   x^2(-24+36x^2-12x^4+x^6),
   \label{d4rsp}\\
   P_{D_{5}}^{{\bf S},\,\bar{\bf S}}(x)&\!\!=\!\!&
   25-3400x^2+13900x^4-20200x^6+12730x^8-3880x^{10}\nonumber\\
   &&\quad +580x^{12}-40x^{14}+x^{16},\\
   P_{D_{6}}^{{\bf S},\,\bar{\bf S}}(x)&\!\!=\!\!&2^{-16}
   \left(951356390625-24582413628000x^2+229552540380000x^4\right.\nonumber\\
   &&\quad -1001859665040000x^6+2271780895320000x^8-2992279237056000x^{10}
   \nonumber\\
   &&\quad +2465846485977600x^{12}-1332743493888000x^{14}+486926396352000x^{16}
   \nonumber\\
   &&\quad -122431951872000x^{18}+21351239884800x^{20}-2577889198080x^{22}
   \nonumber\\
   &&\quad +212745830400x^{24}-11668684800x^{26}+403046400x^{28}-7864320x^{30}
   \nonumber\\
   &&\quad\left. +65536x^{32}\right).
\end{eqnarray}
The equality of the three polynomials for ${\bf V}$, ${\bf S}$ and
$\bar{\bf S}$ in $D_4$, (\ref{d4rsp}) reflects the three fold symmetry of
the $D_4$ Dynkin diagram.

%%%%%%%%%%%%%%%%%%%%%%%%%%%%
%  4.2.5                   %
%%%%%%%%%%%%%%%%%%%%%%%%%%%%
\subsubsection{${\cal R}={\Delta_L}$ for $B_r$ and $D_r$}
\label{brrdell}

The set of long roots of $B_r$ is $\Delta_L=\{\pm({\bf e}_j-{\bf e}_l),
\pm({\bf e}_j+{\bf e}_l)\,|\,1\leq j<l\leq r\}$.
The polynomial $P_{B_r}^{\Delta_L}(k|x)$ can be expressed neatly in terms of
the coefficients of the polynomial $P_{B_r}^{\Delta_S}(k|x)$ (\ref{lagpoly}).
Suppose we have two polynomials in $y$:
\begin{eqnarray}
    &&f=\prod_{i=1}^n(y-x_i^2)=\sum_{i=0}^n(-1)^ia_i\,y^{n-i},\\
    &&g=\prod_{1\leq i<j\leq n}
    \Bigl(y-(x_i-x_j)^2\Bigr)\Bigl(y-(x_i+x_j)^2\Bigr).
\end{eqnarray}
Let us denote $b_i=x_i^2$, then we obtain $g$ as a symmetric polynomial
in $b_i$:
\begin{equation}
   g=\prod_{1\leq i<j\leq n}\Bigl(y^2-2(b_i+b_j)y+(b_i-b_j)^2\Bigr),
\end{equation}
and $\{a_i\}$ are the basis of the  symmetric polynomials in $b_i$:
\begin{equation}
   a_i=\sum_{1\leq j_1<\cdots<j_i\leq n}b_{j_1}\cdots b_{j_i}.
\end{equation}
Thus $g$ can be expressed in terms of the coefficients $\{a_i\}$ of
$f$ with integer coefficients. For example:
\begin{eqnarray}
    n=2:&&g=y^2-2a_1y+a_1^2-4a_2,\label{rfgrel1}\\
    n=3:&&g=y^6-4a_1y^5+2(3a_1^2-a_2)y^4-2(2a_1^3-a_1a_2-13a_3)y^3\nonumber\\
    &&\phantom{g=}
    +(a_1^4+2a_1^2a_2-7a_2^2-24a_1a_3)y^2-2(a_1^2-3a_2)(a_1a_2-9a_3)y
    \nonumber\\
    &&\phantom{g=}
    +a_1^2a_2^2-4a_2^3-4a_1^3a_3+18a_1a_2a_3-27a_3^2.
    \label{rfgrel2}
\end{eqnarray}
If $f$ is of rational coefficients, so is $g$.

We list $P_{B_r}^{\Delta_L}(k|x)$ for lower members of $r$. This includes
$P_{D_r}^{\Delta}(x)$ as a special case of $k=0$. As remarked before,
they are presented as polynomials in $y\equiv x^2$:
\begin{eqnarray}
   P_{B_2}^{\Delta_L}(k|x)&\!\!=\!\!&4(1+k)-4(1+k)y+y^2,\\
   P_{B_3}^{\Delta_L}(k|x)&\!\!=\!\!&
   108\left( 1 + k \right){\left( 2 + k \right) }^2 -
   324\left( 1 + k \right){\left( 2 + k \right)}^2\,y +
   9\,{\left( 2 + k \right) }^2\left( 41 + 32\,k \right)
   y^2\nonumber\\
   && - 4\,\left( 2 + k \right)
   \left( 99 + 88\,k + 16\,k^2 \right) \,y^3 +
   6\left( 2 + k \right)\left( 17 + 8\,k \right) y^4\nonumber\\
   && - 12\,\left( 2 + k \right) \,y^5 + y^6,\\
   P_{B_4}^{\Delta_L}(k|x)&\!\!=\!\!&
   27648\left( 1 + k \right){\left( 2 + k \right) }^2
   {\left( 3 + k \right) }^3 -
   165888\left( 1 + k \right){\left( 2 + k \right) }^2
   {\left( 3 + k \right) }^3\,y\nonumber\\
   && + 4608{\left( 2 + k \right) }^2
   {\left( 3 + k \right) }^3\left( 91 + 82\,k \right) y^2\nonumber\\
   && - 512\left( 2 + k \right){\left( 3 + k \right)}^3
   \left( 2282 + 2777\,k + 792\,k^2 \right)y^3\nonumber\\
   && + 192\left( 2 + k \right){\left( 3 + k \right) }^2%
   \left( 15462 + 20235\,k + 8336\,k^2 + 1088\,k^3 \right)%
   y^4\nonumber\\
   && - 768\left( 2 + k \right)%
   {\left( 3 + k \right) }^2%
   \left( 2085 + 2167\,k + 688\,k^2 + 64\,k^3 \right)y^5\nonumber\\
   && + 64{\left( 3 + k \right) }^2%
   \left( 17634 + 22113\,k + 9480\,k^2 + 1536\,k^3 +
   64\,k^4 \right)y^6\nonumber\\
   && - 768{\left( 3 + k \right) }^2%
   \left( 342 + 327\,k + 96\,k^2 + 8\,k^3 \right)y^7\nonumber\\
   && + 48\left( 3 + k \right)%
   \left( 2514 + 2465\,k + 784\,k^2 + 80\,k^3 \right)y^8\\
   && - 64\left( 3 + k \right)%
   \left( 186 + 123\,k + 20\,k^2 \right)y^9 +
   240{\left( 3 + k \right) }^2y^{10} -
   24\left( 3 + k \right)y^{11} + y^{12}.\nonumber
\end{eqnarray}
As remarked above, $P_{B_r}^{\Delta_L}(k|x)$ is a polynomial in $y$ and
in $k$ with {\em all integer coefficients\/} and is {\em monic\/} in $y$.
The explicit forms of the polynomials $P_{B_r}^{\Delta_L}(k|x)$ ($r=5,6$)
and $P_{D_r}^{\Delta}(x)$ ($r=4,5,6$) can be found in \cite{poly}.
The Dynkin diagram folding $D_{r+1}\to B_r$ relates the polynomials
\begin{equation}
   P_{B_r}^{\Delta_L}(2|x)\left(P_{B_r}^{\Delta_S}(2|x)\right)^2=
   P_{D_{r+1}}^{\Delta}(x)=P_{B_{r+1}}^{\Delta_L}(0|x),
\label{tdrbrredV}
\end{equation}
which is the  root version of the identity (\ref{drbrredV}).

%%%%%%%%%%%%%%%%%%%%%%%%%%%%
%                          %
%%%%%%%%%%%%%%%%%%%%%%%%%%%%
\bigskip
Next we discuss the systems based on the exceptional root systems. For these
we have relied on numerical evaluation of the equilibrium points by 
Mathematica.
Large enough digits of precision is maintained in internal computations,
{\it e.g.}, we keep 2048 digits for $E_8$ Sutherland system.
We have verified in each case that the fit of the polynomial with rational
coefficients gives no detectable errors within the working precision.

%%%%%%%%%%%%%%%%%%%%%%%%%%%%
%  4.3                     %
%%%%%%%%%%%%%%%%%%%%%%%%%%%%
\subsection{$E_r$}

The $E$ series of the root systems, $E_6$, $E_7$ and $E_8$ are simply laced.
The corresponding polynomials do not contain any coupling constants.

%%%%%%%%%%%%%%%%%%%%%%%%%%%%
%  4.3.1                   %
%%%%%%%%%%%%%%%%%%%%%%%%%%%%
\subsubsection{${\cal R}={\bf 27}$ and $\Delta$ for $E_6$}

Polynomials for ${\bf 27}$ and $\Delta$,
\begin{eqnarray}
   P_{E_6}^{\bf 27}(x)&\!\!=\!\!&
   \prod_{\mu\in{\bf 27}}(x-\mu\cdot\tilde{q})\quad(\mu^2=4/3,\;\rho^2=2),
   \label{e6r27o}\\
   P_{E_6}^{\Delta}(x)&\!\!=\!\!&
   \prod_{\rho\in{\Delta}}\left(x-\rho\cdot\tilde{q}\right)\quad (\rho^2=2),
   \label{e6rrseco}
\end{eqnarray}
are slightly simplified for a different normalisation of $\mu\in{\cal R}$:
\begin{eqnarray}
   &&P_{E_6}^{\sqrt{1/3}\,{\bf 27}}(x)=3^{-27/2}P_{E_6}^{\bf 27}(\sqrt{3}x)=
   \prod_{\mu\in{\bf 27}}(x-\hat{\mu}\cdot\tilde{q})\quad
   (\hat{\mu}=\mu/\sqrt{3},\;\hat{\mu}^2=4/9)\nonumber\\
   &&=
   x^3\,\left( 200 - 3600\,x^2 + 24600\,x^4 - 83980\,x^6 +
   162945\,x^8 - 192840\,x^{10}  \right.\nonumber\\
   &&\quad\quad \left.  + 144876\,x^{12}-
   70416\,x^{14} + 22170\,x^{16} - 4440\,x^{18} +
   540\,x^{20} - 36\,x^{22} + x^{24} \right),
   \label{e6r27}\\
   &&P_{E_6}^{\sqrt{1/3}\,\Delta}(x)=3^{-36}P_{E_6}^{\Delta}(\sqrt{3}x)=
   \prod_{\rho\in{\Delta}}\left(x-\hat{\rho}\cdot\tilde{q}\right)\quad
   (\hat{\rho}=\rho/\sqrt{3},\;\hat{\rho}^2=2/3)\nonumber\\
   &&=\Bigl( 81920 - 1474560\,x^2 + 8970240\,x^4 - 22749184\,x^6 +
   28505088\,x^8 - 19829760\,x^{10} \nonumber\\
   &&\qquad + 8239872\,x^{12} - 2128896\,x^{14} +
   346944\,x^{16} - 35328\,x^{18} + 2160\,x^{20} - 72\,x^{22} + x^{24}
   \Bigr) \nonumber\\
   &&\phantom{=}\times \Bigl( 200 - 3600\,x^2 + 24600\,x^4 - 83980\,x^6 +
   162945\,x^8 - 192840\,x^{10} + 144876\,x^{12}\nonumber\\
   &&\qquad - 70416\,x^{14} +
   22170\,x^{16} - 4440\,x^{18} + 540\,x^{20} - 36\,x^{22} + x^{24} \Bigr)^2
   \label{e6rrsec}
\end{eqnarray}
It is interesting to note that the second factor of $P_{E_6}^{\Delta}(x)$,
(\ref{e6rrsec}), is the same as $P_{E_6}^{\bf 27}(x)/x^3$,
which is the same polynomial appearing in (\ref{e6r27}) and (\ref{efredr272}).
Again it should be stressed that these polynomials are {\em monic\/}
and all the coefficients are {\em integers\/}.

%%%%%%%%%%%%%%%%%%%%%%%%%%%%
%  4.3.2                   %
%%%%%%%%%%%%%%%%%%%%%%%%%%%%
\subsubsection{${\cal R}={\bf 56}$ for $E_7$}

Polynomial for ${\bf 56}$,
\begin{equation}
   P_{E_7}^{\bf 56}(x)=
   \prod_{\mu\in{\bf 56}}(x-\mu\cdot\tilde{q})\quad(\mu^2=3/2,\;\rho^2=2),
   \label{e7r56o}
\end{equation}
is slightly simplified for a different normalisation of $\mu$:
\begin{eqnarray}
   &&P_{E_7}^{\sqrt{2}\,{\bf 56}}(x)=2^{28}P_{E_7}^{\bf 56}(x/\sqrt{2})=
   \prod_{\mu\in{\bf 56}}(x-\hat{\mu}\cdot\tilde{q})\quad
   (\hat{\mu}=\sqrt{2}\mu,\;\hat{\mu}^2=3)\nonumber\\
   &&=2044117922661550386613265625 -
    48583441852490416903125286500\,x^2\nonumber\\
   &&\phantom{=} +
    403943437764362721049483097250\,x^4 -
    1594876299784237542505579618500\,x^6\nonumber\\
   &&\phantom{=} +
    3423181532874686547792360316875\,x^8 -
    4470973846715160163197028791000\,x^{10}\nonumber\\
   &&\phantom{=} +
    3844463042762881314328636794900x^{12} -
    2298706753677324429083230164600x^{14}\nonumber\\
   &&\phantom{=} +
    994190889968661674517540390225\,x^{16} -
    320292296385170629680242995500\,x^{18}\nonumber\\
   &&\phantom{=} +
    78600569652362205629789205150\,x^{20} -
    14948636823173617875192068460\,x^{22}\nonumber\\
   &&\phantom{=} +
    2232949785098933644991402715\,x^{24} -
    264680665744227895592493840\,x^{26}\nonumber\\
   &&\phantom{=} +
    25089285771398909108223000\,x^{28} -
    1912398423761929885120080\,x^{30}\nonumber\\
   &&\phantom{=} +
    117632735062147883037411\,x^{32} -
    5848529412061451267964\,x^{34}\nonumber\\
   &&\phantom{=} +
    234966118304680273854\,x^{36} -
    7609794291104570460\,x^{38}\nonumber\\
   &&\phantom{=} +
    197734877929087065\,x^{40} -
    4090765650038424\,x^{42} + 66612822142356\,x^{44}\nonumber\\
   &&\phantom{=} -
    839599815096\,x^{46} + 7991799795\,x^{48} -
    55327860\,x^{50}\nonumber\\
   &&\phantom{=} + 261954\,x^{52} - 756\,x^{54} + x^{56}
   \label{e7rat56}
\end{eqnarray}

%%%%%%%%%%%%%%%%%%%%%%%%%%%%
%  4.3.3                   %
%%%%%%%%%%%%%%%%%%%%%%%%%%%%
\subsubsection{${\cal R}={\Delta}$ for $E_7$ and $E_8$}

The polynomials $P_{E_7}^{\Delta}(x)$ and $P_{E_8}^{\Delta}(x)$ are too
long to be displayed here. See \cite{poly} for explicit forms.
It should be stressed that five {\em monic\/} polynomials in $x$,
$P_{E_6}^{\sqrt{1/3}\,{\bf 27}}(x)$ (\ref{e6r27}) (and $P_{E_6}^{\bf 27}(x)$
(\ref{e6r27o})), $P_{E_6}^{\Delta}(x)$ (\ref{e6rrsec}),
$P_{E_7}^{\sqrt{2}\,\bf 56}(x)$ (\ref{e7rat56}), $P_{E_7}^{\Delta}(x)$ and
$P_{E_8}^{\Delta}(x)$ have {\em integer coefficients only\/}.

%%%%%%%%%%%%%%%%%%%%%%%%%%%%
%  4.4                     %
%%%%%%%%%%%%%%%%%%%%%%%%%%%%
\subsection{$F_4$}

The theory has two coupling constants $g_L$ and $g_S$ for the long
($\rho_L^2=2$) and short ($\rho_S^2=1$) roots.
We present the polynomials as a function of $k\equiv g_S/g_L$.

%%%%%%%%%%%%%%%%%%%%%%%%%%%%
%  4.4.1                   %
%%%%%%%%%%%%%%%%%%%%%%%%%%%%
\subsubsection{${\cal R}=\Delta_L$ for $F_4$}

\begin{eqnarray}
   P_4^L(k|y)&\!\!\equiv\!\!&P_{F_4}^{\Delta_L}(k|x)=
   \prod_{\rho\in{\Delta_{L}}}\left(x-\rho\cdot\tilde{q}\right)
   =\prod_{\rho\in{\Delta_{L+}}}\left(y-(\rho\cdot\tilde{q})^2\right)
   \quad (\rho_L^2=2) \nonumber\\
   &\!\!=\!\!&746496(1 + k)^6(2 + k)^2(1 + 2k) -
   4478976(1 + k)^6(2 + k)^2(1 + 2k)y\nonumber\\
   && + 124416(1 + k)^5(2 + k)^2(1 + 2k)(91 + 64\,k)y^2\nonumber\\
   && -
   13824(1 + k)^5(2 + k)\left(2282 + 6049k + 3712\,k^2 + 512\,k^3\right)y^3
   \nonumber\\
   && +
   15552(1 + k)^4(2 + k)\left(1718 + 5027k + 4288\,k^2 + 1024\,k^3\right)y^4
   \nonumber\\
   && -
   20736(1 + k)^4(2 + k)\left(695 + 1472k + 704\,k^2\right)y^5\nonumber\\
   && +
   1728(1 + k)^3\left(5878 + 16235k + 14408k^2 + 4096\,k^3\right)y^6
   \nonumber\\
   && - 62208(1 + k)^3(38 + 71k + 32k^2)y^7 +
   432(1 + k)^2(838 + 1627k + 784k^2)y^8\nonumber\\
   && - 576(1 + k)^2(62 + 61k)y^9 + 2160(1 + k)^2y^{10} - 72(1 + k)y^{11}
   + y^{12}.
   \label{f4r24l}
\end{eqnarray}

%%%%%%%%%%%%%%%%%%%%%%%%%%%%
%  4.4.2                   %
%%%%%%%%%%%%%%%%%%%%%%%%%%%%
\subsubsection{${\cal R}=\Delta_S$ for $F_4$}

\begin{eqnarray}
   P_4^S(k|y)&\!\!\equiv\!\!&P_{F_4}^{\Delta_S}(k|x)=
   \prod_{\rho\in{\Delta_{S}}}\left(x-\rho\cdot\tilde{q}\right)
   =\prod_{\rho\in{\Delta_{S+}}}\left(y-(\rho\cdot\tilde{q})^2\right)
   \quad (\rho_S^2=1)
   \nonumber\\
   &\!\!=\!\!&729k^3(1 + k)^6(2 + k)(1 + 2k)^2/4 -
   2187k^2(1 + k)^6(2 + k)(1 + 2k)^2y\nonumber\\
   && +
   243k(1 + k)^5(2 + k)(1 + 2k)^2(64 + 91k)/2\,y^2\nonumber\\
   && -
   27(1 + k)^5(1 + 2k)\left(512 + 3712k + 6049k^2 + 2282k^3\right)y^3
   \nonumber\\
   && + 243(1 + k)^4(1 + 2k)(1024 + 4288k + 5027k^2 +
   1718k^3)/4\,y^4\nonumber\\
   && - 162(1 + k)^4(1 + 2k)\left(704 + 1472k + 695k^2\right)y^5\nonumber\\
   && + 27(1 + k)^3\left(4096 + 14408k + 16235k^2 + 5878k^3\right)y^6
   \nonumber\\
   && - 1944(1 + k)^3\left(32 + 71k + 38k^2\right)y^7 +
   27(1 + k)^2\left(784 + 1627k + 838k^2\right)y^8\nonumber\\
   && - 72(1 + k)^2(61 + 62k)
   y^9 + 540(1 + k)^2y^{10} - 36(1 + k)y^{11} + y^{12}.
\end{eqnarray}
They are related with each other reflecting the self-duality of the $F_4$
root system. If one replaces $k$ by $1/k$ and $y$ by $y/(2k)$ in
$P_{4}^{S}(k|y)$, one obtains $P_{4}^{L}(k|y)/(2k)^{12}$:
\begin{equation}
   P_{4}^{L}(k|y)=(2k)^{12}P_{4}^{S}(1/k|y/2k),\quad
   \mbox{or}\quad
   P_{F_4}^{\Delta_L}(k|x)=(2k)^{12}P_{F_4}^{\Delta_S}(1/k|x/\sqrt{2k}).
\end{equation}
It is well-known that $F_4$ with the coupling ratio $k=g_S/g_L=2$ is
obtained from $E_6$ by folding. This relates $F_4$ polynomials to $E_6$
polynomials:
\begin{equation}
   P_{F_4}^{\Delta_S}(2|x)=P_{E_6}^{\bf 27}(x)/x^3,\quad
   P_{F_4}^{\Delta_L}(2|x)\left(P_{F_4}^{\Delta_S}(2|x)\right)^2=
   P_{E_6}^{\Delta}(x).
   \label{efredr272}
\end{equation}
Both of them have trigonometric counterparts as will be shown later
(\ref{efredt272})-(\ref{efredt274}).
The two polynomials $P_{F_4}^{\Delta_L}(k|x)$ and
$P_{F_4}^{\sqrt{2}\Delta_S}(k|x)$ have {\em integer coefficients only\/}.
This property seems to be  inherited from $E_6$, too.

%%%%%%%%%%%%%%%%%%%%%%%%%%%%
%  4.5                     %
%%%%%%%%%%%%%%%%%%%%%%%%%%%%
\subsection{$G_2$}

The theory has two coupling constants $g_L$ and $g_S$ for the long
($\rho_L^2=2$) and short ($\rho_S^2=2/3$) roots.
We present the polynomials as a function of $k\equiv g_S/g_L$.

%%%%%%%%%%%%%%%%%%%%%%%%%%%%
%  4.5.1                   %
%%%%%%%%%%%%%%%%%%%%%%%%%%%%
\subsubsection{${\cal R}=\Delta_L$ for $G_2$}

\begin{eqnarray}
   P_2^L(k|y)&\!\!\equiv\!\!&P_{G_2}^{\Delta_L}(k|x)=
   \prod_{\rho\in{\Delta_{L}}}\left(x-\rho\cdot\tilde{q}\right)
   =\prod_{\rho\in{\Delta_{L+}}}\left(y-(\rho\cdot\tilde{q})^2\right)
   \quad (\rho_L^2=2) \nonumber\\
   &\!\!=\!\!&-27(1 + k)^2/2 + 81(1 + k)^2/4\,y - 9(1 + k)y^2 + y^3.
\end{eqnarray}

%%%%%%%%%%%%%%%%%%%%%%%%%%%%
%  4.5.2                   %
%%%%%%%%%%%%%%%%%%%%%%%%%%%%
\subsubsection{${\cal R}=\Delta_S$ for $G_2$}

\begin{eqnarray}
   P_2^S(k|y)&\!\!\equiv\!\!&P_{G_2}^{\Delta_S}(k|x)=
   \prod_{\rho\in{\Delta_{S}}}\left(x-\rho\cdot\tilde{q}\right)
   =\prod_{\rho\in{\Delta_{S+}}}\left(y-(\rho\cdot\tilde{q})^2\right)
   \quad (\rho_S^2=2/3) \nonumber\\
   &\!\!=\!\!& -k(1 + k)^2/2 + 9(1 + k)^2/4\,y - 3(1 + k)y^2 + y^3.
   \label{grsr}
\end{eqnarray}
They are related with each other reflecting the self-duality of
the $G_2$ root system:
\begin{equation}
   P_{2}^{L}(k|y)=(3k)^3P_{2}^{S}(1/k|y/3k),\quad \mbox{or}\quad
   P_{G_2}^{\Delta_L}(k|x)=(3k)^3P_{G_2}^{\Delta_S}(1/k|x/\sqrt{3k}).
\end{equation}
The $G_2$ Calogero system with the coupling ratio $k=g_S/g_L=3$
is obtained from that of $D_4$ by the three-fold folding $D_4\to G_2$.
This implies  analogous relations to (\ref{efredr272})
\begin{equation}
   P_{G_2}^{\Delta_S}(3|x)=P_{D_4}^{{\cal R}}(x)/x^2\quad
   ({\cal R}={\bf V}, {\bf S}, {\bf \bar{S}}),\quad
   P_{G_2}^{\Delta_L}(3|x)\left(P_{G_2}^{\Delta_S}(3|x)\right)^3=
   P_{D_4}^{\Delta}(x).
   \label{ratd4g2}
\end{equation}
Both of them have trigonometric counterparts, too, as will be shown later.
The two polynomials $P_{G_2}^{\sqrt{2}\Delta_L}(k|x)$ and
$P_{G_2}^{\sqrt{2}\Delta_S}(k|x)$ have {\em integer coefficients only\/}.
This property seems to be  inherited from $D_4$.

%%%%%%%%%%%%%%%%%%%%%%%%%%%%
%                          %
%%%%%%%%%%%%%%%%%%%%%%%%%%%%
\bigskip
Thirdly let us discuss the systems based on non-crystallographic root systems.

%%%%%%%%%%%%%%%%%%%%%%%%%%%%
%  4.6                     %
%%%%%%%%%%%%%%%%%%%%%%%%%%%%
\subsection{$I_2(m)$}

The equilibrium points are easily obtained when parametrised by
the two-dimensional polar coordinates \cite{cs}:
\begin{equation}
   \bar{q}=(\bar{q}_1,\bar{q}_2)=\bar{r}(\sin\bar{\varphi},\cos\bar{\varphi}),
   \label{polar}
\end{equation}
\begin{equation}
   \bar{r}^2={mg\over{\omega}},\;\;
   \bar{\varphi}={\pi\over{2m}}\;\; (m:\ \mbox{odd})\;;
   \quad
   \bar{r}^2={m(g_e+g_o)\over{2\omega}},\;\;
   \tan{m\bar{\varphi}\over2}=\sqrt{g_e\over{g_o}}\;\; (m:\ \mbox{even}),
\end{equation}
in which $g$ is the coupling constant in the simply laced odd $m$ theory,
whereas $g_o$ ($g_e$) is the coupling constant for odd (even) roots in
the non-simply laced even $m$ theory.
As ${\cal R}$ we choose the set of the vertices of the regular $m$-gon
$R_{m}$ on which the dihedral group $I_2(m)$ acts:
\begin{eqnarray}
   &&R_m=\Bigl\{(\cos({2j\pi/ m}+t_0), \sin({2j\pi/ m}+t_0) )
   \in{\mathbb R}^2\Bigm|\ j=1,\ldots,m\Bigr\},
   \label{rmvert}\\
   &&t_0=\pi/2m\;\;(m:\ \mbox{odd})\;;\quad
   t_0=0\;\;(m:\ \mbox{even}). \nonumber
\end{eqnarray}
The polynomial $\prod_{\mu\in{R_m}}(x-\mu\cdot\tilde{q})$ (\ref{calpolydef})
is obtained trivially:
\begin{equation}
   P_m(x)\equiv P_{I_2(m)}^{R_m}(x)=
   \prod_{\mu\in{R_m}}(x-\mu\cdot\tilde{q})
   =\prod_{j=1}^m
   \left(x-\sin\Bigl({2j\pi\over m}+{\varphi_0\over{m}}\Bigr)\right),
   \label{i2poly1}
\end{equation}
in which
\begin{equation}
   \varphi_0=\pi\;\;(m:\ \mbox{odd})\;;\quad
   \varphi_0=2\arctan\sqrt{k},\;\; k\equiv g_e/{g_o}\;\; (m:\ \mbox{even}).
\end{equation}
For odd $m$ $P_m(x)$ is proportional to the Chebyshev polynomial 
of the first kind $T_m(x)$
(see (\ref{cheby})). For even  $m$ and for the equal coupling $g_e=g_o$,
$P_m(x)$ is also proportional to the Chebyshev polynomial $T_m(x)$ and thus the
entire $\{P_m(x)=2^{1-m}T_m(x)\}$ constitute orthogonal polynomials \cite{cs}.
For generic coupling $g_e\neq g_o$ the orthogonality no longer holds.
This can be seen most easily by the explicit forms of the lower members
of $P_{even}$ in the non-singular limiting cases, $g_e=0$ and $g_o=0$:
\begin{eqnarray}
   g_e=0&:&\;\; x^2,\;\; x^2(x^2-1),\;\; x^2(x^2-3/4)^2,
   \;\; x^2(x^2-1/2)^2(x^2-1),\;\ldots,\nonumber\\
   g_o=0&:&\;\; x^2-1,\;\; (x^2-1/2)^2,\;\; (x^2-1)(x^2-1/4)^2,\;\; 
   (x^4-x^2+1/8)^2,\;\ldots,
\end{eqnarray}
which have definite sign in $-1<x<1$.

The following equivalences are well-known: $A_2\equiv I_2(3)$,
$B_2\equiv I_2(4)$ and $G_2\equiv I_2(6)$. The $I_2(3)$ polynomial
corresponds to the $A_2$ polynomial of vector ${\bf V}$,
\begin{equation}
   P_{I_2(3)}^{R_3}(x)=\frac14T_3(x)
   =\frac{1}{16\sqrt{2}}H_3(\sqrt{2}x)=P_{A_2}^{{\bf V}/\sqrt{2}}(x).
\end{equation}
As for $I_2(4)$ polynomial, we obtain from (\ref{i2poly1})
\begin{equation}
   P_{I_2(4)}^{R_4}(x)=x^4-x^2 +{k\over{4(1+k)}},\quad k\equiv g_e/g_o.
\end{equation}
For the $B_2$ system,
the Laguerre polynomial with $\alpha=k-1\equiv g_e/g_o-1$ reads
\[
   L_2^{(\alpha)}(y)={1\over2}y^2-(k+1)y+{k(1+k)/2},\quad \alpha=k-1.
\]
They are proportional to each other upon identification $y=2(1+k)x^2$.
The $I_2(6)$ polynomial obtained from (\ref{i2poly1}) reads,
after some calculation
\begin{equation}
   P_{I_2(6)}^{R_6}(x)=x^6-{3\over2}x^4+{9\over{16}}x^2-
   {k\over{16(1+k)}},\quad k=g_e/g_o,
\end{equation}
which is proportional to $P_{2}^{S}(k|y)$ (\ref{grsr}) upon the same
identification as above $y=2(1+k)x^2$.

%%%%%%%%%%%%%%%%%%%%%%%%%%%%
%  4.7                     %
%%%%%%%%%%%%%%%%%%%%%%%%%%%%
\subsection{$H_3$ and $H_4$}

The non-crystallographic $H_3$ and $H_4$ are  simply laced root systems.
In both cases the roots are normalised to 2, as with the other simply
laced root systems, $\rho^2=2$.
Then both monic  polynomials $P_{H_3}^{\Delta}(x)$ and
$P_{H_4}^{\Delta}(x)$ have {\em integer coefficients only\/}.

%%%%%%%%%%%%%%%%%%%%%%%%%%%%
%  4.7.1                   %
%%%%%%%%%%%%%%%%%%%%%%%%%%%%
\subsubsection{${\cal R}=\Delta$ for $H_3$}

\begin{eqnarray}
   P_{3}^{\Delta}(y)&\!\!\equiv\!\!&P_{H_3}^{\Delta}(x)=
   \prod_{\rho\in{\Delta}}\left(x-\rho\cdot\tilde{q}\right)
   =\prod_{\rho\in{\Delta_{+}}}\left(y-(\rho\cdot\tilde{q})^2\right)
   \quad (\rho^2=2)\nonumber\\
   &\!\!=\!\!&\left( -450 + 225\,y - 30\,y^2 + y^3 \right)\nonumber\\
   &&\times
    \left( 5625 - 22500\,y + 27000\,y^2 - 9600\,y^3 +
    1200\,y^4 - 60\,y^5 + y^6 \right)\nonumber\\
   && \times
    \left( 22500 - 67500\,y + 46125\,y^2 - 11700\,y^3 +
    1275\,y^4 - 60\,y^5 + y^6 \right).
\end{eqnarray}

%%%%%%%%%%%%%%%%%%%%%%%%%%%%
%  4.7.2                   %
%%%%%%%%%%%%%%%%%%%%%%%%%%%%
\subsubsection{${\cal R}=\Delta$ for $H_4$}

\begin{eqnarray}
   &&\hspace{-11mm}\;\;P_{4}^{\Delta}(y)\equiv P_{H_4}^{\Delta}(x)=
   \prod_{\rho\in{\Delta}}\left(x-\rho\cdot\tilde{q}\right)
   =\prod_{\rho\in{\Delta_{+}}}\left(y-(\rho\cdot\tilde{q})^2\right)
   \quad (\rho^2=2)\nonumber\\
   &&\hspace{-11mm}=
   \bigl( 656100000000 - 1093500000000\,y + 601425000000\,y^2 -
      154305000000\,y^3 + 21343500000\,y^4 \nonumber\\
   &&\hspace{-11mm}\;\;- 1701000000\,y^5 + 80392500\,y^6 -
      2250000\,y^7 + 36000\,y^8 - 300\,y^9 + y^{10} \bigr) \,\nonumber\\
   &&\hspace{-11mm}\times
   \bigl( 747338906250000000000 - 9964518750000000000000\,y +
      45172485000000000000000\,y^2 \nonumber\\
   &&\hspace{-11mm}\;\;- 90926233593750000000000\,y^3 +
      92928548278125000000000\,y^4 \nonumber\\
   &&\hspace{-11mm}\;\;- 52841916742500000000000\,y^5 +
      18358385767875000000000\,y^6 \nonumber\\
   &&\hspace{-11mm}\;\;- 4169745135000000000000\,y^7 +
      648844128590625000000\,y^8 - 71483472810000000000\,y^9 \nonumber\\
   &&\hspace{-11mm}\;\;
   + 5707114499700000000\,y^{10} - 335580296625000000\,y^{11} +
      14683267406250000\,y^{12} \nonumber\\
   &&\hspace{-11mm}\;\;- 480384270000000\,y^{13} +
      11739694500000\,y^{14} - 212600700000\,y^{15} + 2804085000\,y^{16}
   \nonumber\\
   &&\hspace{-11mm}\;\;
   - 26100000\,y^{17} + 162000\,y^{18} - 600\,y^{19} + y^{20} \bigr) \,
   \nonumber\\
   &&\hspace{-11mm}\times\bigl( 1362025156640625000000000000000000 -
      20430377349609375000000000000000000\,y \nonumber\\
   &&\hspace{-11mm}\;\;+ 110664543977050781250000000000000000\,y^2 -
      280672524028933593750000000000000000\,y^3 \nonumber\\
   &&\hspace{-11mm}\;\;+ 406550434997274609375000000000000000\,y^4 -
      377089903500479578125000000000000000\,y^5 \nonumber\\
   &&\hspace{-11mm}\;\;+ 240385775914964970703125000000000000\,y^6 -
      110467977515351929687500000000000000\,y^7 \nonumber\\
   &&\hspace{-11mm}\;\;+ 37879740110299305937500000000000000\,y^8 -
      9947615045119592062500000000000000\,y^9 \nonumber\\
   &&\hspace{-11mm}\;\;+ 2041289604408317542031250000000000\,y^{10} -
      332506194678726581250000000000000\,y^{11} \nonumber\\
   &&\hspace{-11mm}\;\;+ 43529340095868749062500000000000\,y^{12} -
      4624592400554729343750000000000\,y^{13} \nonumber\\
   &&\hspace{-11mm}\;\;+ 401746375286214215625000000000\,y^{14} -
      28701181376029878750000000000\,y^{15} \nonumber\\
   &&\hspace{-11mm}\;\;+ 1693173350921514750000000000\,y^{16} -
      82699244991680625000000000\,y^{17} \nonumber\\
   &&\hspace{-11mm}\;\;+ 3348318244893890625000000\,y^{18} -
      112349936407545000000000\,y^{19} \nonumber\\
   &&\hspace{-11mm}\;\;+ 3118565868993450000000\,y^{20} -
      71352951283125000000\,y^{21} + 1337980766062500000\,y^{22} \nonumber\\
   &&\hspace{-11mm}\;\;- 20388872475000000\,y^{23} + 249452622000000\,y^{24} -
      2408494500000\,y^{25} + 17897422500\,y^{26} \nonumber\\
   &&\hspace{-11mm}\;\;- 98550000\,y^{27} +
      378000\,y^{28} - 900\,y^{29} + y^{30} \bigr).
\end{eqnarray}

%%%%%%%%%%%%%%%%%%%%%%%%%%%%%%%%%%%%%%%%%%%%%%%%%%%%%%%%%%%%%%%
%                                                             %
%  5. Sutherland Systems                                      %
%                                                             %
%%%%%%%%%%%%%%%%%%%%%%%%%%%%%%%%%%%%%%%%%%%%%%%%%%%%%%%%%%%%%%%
\section{Sutherland Systems}
\label{csutsys}
\setcounter{equation}{0}

Let us first discuss the systems based on the classical root systems.
%%%%%%%%%%%%%%%%%%%%%%%%%%%%
%  5.1                     %
%%%%%%%%%%%%%%%%%%%%%%%%%%%%
\subsection{$A_r$}

The equilibrium position is ``{\em equally-spaced\/}" (see eq.(5.14)
of \cite{cs}) and translational invariant.
We choose the constant shift such that the coordinate of ``center of mass"
vanishes, $\displaystyle \sum_{j=1}^{r+1}\bar{q}_j=0$:
\begin{equation}
   \bar{q}_j=\frac{\pi(r+1-j)}{r+1}-\frac{\pi r}{2(r+1)}=
   \frac{\pi}{2}-\frac{\pi(2j-1)}{2(r+1)}=-\bar{q}_{r+2-j}
   \quad(j=1,\ldots,r+1).
   \label{eqspaced}
\end{equation}

%%%%%%%%%%%%%%%%%%%%%%%%%%%%
%  5.1.1                   %
%%%%%%%%%%%%%%%%%%%%%%%%%%%%
\subsubsection{${\cal R}={\bf V}$ for $A_r$}

For the vector weight $\mu_j\in{\bf V}$ (\ref{ArV}), $\mu_j\cdot\bar{q}$ is
independent on the constant shift of $\bar{q}$.
The above choice (\ref{eqspaced}) leads to
\begin{equation}
   \mu_j\cdot\bar{q}={\pi\over2}-{\pi(2j-1)\over{2(r+1)}}=q_j,
   \quad -{\pi\over2}<\mu_j\cdot\bar{q}<{\pi\over2}\quad(j=1,\ldots,r+1).
   \label{Armujq}
\end{equation}
In this case the polynomial (\ref{sutpolysdef}) is given by
\begin{equation}
   P_r^{\bf V}(x)\equiv P_{A_r,\,s}^{\bf V}(x)
   =\prod_{j=1}^{r+1}\Bigl(x-\sin(\mu_j\cdot\bar{q})\Bigr)
   =\prod_{j=1}^{r+1}\left(x-\cos{\pi(2j-1)\over{2(r+1)}}\right)
   =2^{-r}T_{r+1}(x)\,.
   \label{archev}
\end{equation}
Here $T_n(\cos\varphi)=\cos(n\varphi)$ is the Chebyshev polynomial
of the first kind, whose Rodrigues' formula is
\begin{equation}
   T_n(x)=\frac{(-1)^n}{(2n-1)!!}(1-x^2)^{1/2}\left({d\over{dx}}\right)^n
   (1-x^2)^{n-1/2}=2^{n-1}x^n+\cdots.
   \label{cheby}
\end{equation}
They are orthogonal to each other:
\begin{equation}
   \int_{-1}^1{P_r^{\bf V}(x)P_s^{\bf V}(x)\over{\sqrt{1-x^2}}}\,dx\,
   \propto\delta_{r\,s}.
\end{equation}
This is a new result. Another definition
\begin{equation}
   P_{A_r,\,s}^{{\bf V}\;{}^\prime}(x)
   =\prod_{j=1}^{r+1}\Bigl(x-2\sin(\mu_j\cdot\bar{q})\Bigr)=
   2T_{r+1}(x/2)=2^{r+1}P_{A_r,\,s}^{\bf V}(x/2)
\end{equation}
provides a {\em monic} polynomial with {\em all integer coefficients\/}.

It is easy to see that
\[
   P_{A_r,\,c}^{\bf V}(x)
   =\prod_{j=1}^{r+1}\Bigl(x-\cos(\mu_j\cdot\bar{q})\Bigr)=
   \prod_{j=1}^{r+1}\left(\vTm x-\sin\frac{\pi(2j-1)}{2(r+1)}\right)
\]
does not give rational polynomials, for example,
$P_{A_1,\,c}^{\bf V}(x)=x^2-\sqrt{2}x+1/2$.
In fact, in most cases the polynomial $P_{\Delta,\,c}^{\cal R}(x)$ is
not of rational coefficients.
In the rest of this paper we will not consider this type of polynomials.

The other polynomials,
\begin{eqnarray*}
   &&P_{A_r,\,s2}^{\bf V}(x)
   =\prod_{j=1}^{r+1}\Bigl(x-\sin(2\mu_j\cdot\bar{q})\Bigr)
   =\prod_{j=1}^{r+1}\left(\vTm x-\sin\frac{\pi(2j-1)}{r+1}\right),\\
   &&P_{A_r,\,c2}^{\bf V}(x)
   =\prod_{j=1}^{r+1}\Bigl(x-\cos(2\mu_j\cdot\bar{q})\Bigr)
   =\prod_{j=1}^{r+1}\left(\vTm x+\cos\frac{\pi(2j-1)}{r+1}\right),
\end{eqnarray*}
are essentially the same as $P_{A_r,\,s}^{\bf V}(x)$, (\ref{archev}).
Only the constant term can be different:
\begin{equation}
   P_{A_r,\,s2}^{\bf V}(x)-P_{A_r,\,s}^{\bf V}(x)=-2^{-r}\sin\frac{\pi r}{2},
   \quad
   P_{A_r,\,c2}^{\bf V}(x)-P_{A_r,\,s}^{\bf V}(x)=(-1)^{r+1}2^{-r}.
\end{equation}
Thus we consider only the polynomial
$\displaystyle P_{A_r,\,s}^{\cal R}(x)=\prod_{\mu\in{\cal R}}
\Bigl(x-\sin(\mu\cdot\bar{q})\Bigr)$ for various ${\cal R}$ of $A_r$.

%%%%%%%%%%%%%%%%%%%%%%%%%%%%
%  5.1.2                   %
%%%%%%%%%%%%%%%%%%%%%%%%%%%%
\subsubsection{${\cal R}={\bf V}_i$ for $A_r$}

{}From (\ref{ArViset}) and (\ref{Armujq}), the polynomial (\ref{sutpolysdef})
is given by
\begin{equation}
   P_{A_r,\,s}^{{\bf V}_i}(x)=\prod_{1\leq j_1<\cdots<j_i\leq r+1}
   \Bigl(x-\sin(\bar{q}_{j_1}+\cdots\bar{q}_{j_i})\Bigr)
   =P_{A_r,\,s}^{{\bf V}_{r+1-i}}(x).
   \label{ArVis}
\end{equation}
This polynomial can be expressed in terms of the coefficients of $T_{r+1}(x)$
by the same method as given in section \ref{DMBCr}, and 
$2^{D_i}P_{A_r,s}^{{\bf V}_i}(x/2)$ gives a monic polynomial with 
integer coefficients.
See \cite{poly} for the explicit forms of the polynomials 
$P_{A_r,\,s}^{{\bf V}_i}(x)$ of lower rank $r$.

As in the Calogero case, we report only on ${\bf V}_2$:
\begin{eqnarray}
   &&\;\;P_{A_r,\,s}^{{\bf V}_2}(x)=\prod_{1\leq j<\,l\,\leq r+1}
   \Bigl(x-\sin(\bar{q}_j+\bar{q}_l)\Bigr)\\
   &&=
   \left\{\begin{array}{ll}
   {\displaystyle x^{(r+1)/2}\prod_{1\leq j<\,l\,\leq(r+1)/2}
   \Bigl(x^2-\sin^2(\bar{q}_j-\bar{q}_l)\Bigr)
   \Bigl(x^2-\sin^2(\bar{q}_j+\bar{q}_l)\Bigr)}&\mbox{$r$ : odd}\\
   {\displaystyle x^{r/2}\prod_{j=1}^{r/2}(x^2-\sin^2\bar{q}_j)
   \;\cdot\!\!\!\!\! \prod_{1\leq j<\,l\,\leq r/2}
   \Bigl(x^2-\sin^2(\bar{q}_j-\bar{q}_l)\Bigr)
   \Bigl(x^2-\sin^2(\bar{q}_j+\bar{q}_l)\Bigr)}&\mbox{$r$ : even}.
   \end{array}\right.\nonumber
\end{eqnarray}
Based on the fact that the zeros of Chebyshev and Jacobi polynomials are
related as seen from the formulae (\ref{ChebJaco1}) and (\ref{ChebJaco2}),
this can be expressed by using the polynomials associated with the $BC_r$
Sutherland systems in the following way:
\begin{eqnarray}
   &&P_{A_{2r-1},\,s}^{{\bf V}_2}(x)
   =2^{-r(r-1)}x^rP_{BC_r,\,c2}^{\Delta_{M+}}(0,1/2|1-2x^2),\\
   &&P_{A_{2r},\,s}^{{\bf V}_2}(x)
   =2^{-r(r-1)}x^{r-1}P_{A_{2r},\,s}^{{\bf V}}(x)
   P_{BC_r,\,c2}^{\Delta_{M+}}(1,1/2|1-2x^2).
\end{eqnarray}
The explicit forms of the functions
$P_{BC_r,\,c2}^{\Delta_{M+}}(k_1,k_2|x)$ for lower $r$ are given
in section \ref{DMBCr}.

%%%%%%%%%%%%%%%%%%%%%%%%%%%%
%  5.1.3                   %
%%%%%%%%%%%%%%%%%%%%%%%%%%%%
\subsubsection{${\cal R}=\Delta$ for $A_r$}

The polynomial has a factorized form:
\begin{equation}
   P_{A_r,\,s}^{\Delta}(x)=
   \prod_{1\leq j<l\leq r+1}\Bigl(x^2-\sin^2(\bar{q}_j-\bar{q}_l)\Bigr)
   =\left\{\begin{array}{ll}
    x^2-1&\;(r=1)\\
    {\displaystyle
    x^{-r-1}P_{A_r,\,s2}^{{\bf V}}(x)\Bigl(P_{A_r,\,s}^{{\bf V}_2}(x)\Bigr)^2}
    &\;(r\geq 2)\,.
   \end{array}\right.
\end{equation}
It is elementary to evaluate $P_{A_r,\,s}^{\Delta}(x)$ for lower rank:
\begin{eqnarray*}
   P_{A_r,\,s}^{\Delta}(x)&\!\!=\!\!&\prod_{1\leq j<l<r+1}
   \left(x^2-\sin^2\Bigl({\pi(l-j)\over{r+1}}\Bigr)\right),\\
   P_{A_1,\,s}^{\Delta}(x)&\!\!=\!\!&x^2-1,\\
   P_{A_2,\,s}^{\Delta}(x)&\!\!=\!\!&2^{-6}(4x^2-3)^3,\\
   P_{A_3,\,s}^{\Delta}(x)&\!\!=\!\!&2^{-4}(x^2-1)^2(2x^2-1)^4,\\
   P_{A_4,\,s}^{\Delta}(x)&\!\!=\!\!&2^{-20}(5-20x^2+16x^4)^5,\\
   P_{A_5,\,s}^{\Delta}(x)&\!\!=\!\!&2^{-24}(x^2-1)^3(4x^2-1)^6(4x^2-3)^6,\\
   P_{A_6,\,s}^{\Delta}(x)&\!\!=\!\!&
   2^{-42}(-7 + 56\,x^2 - 112\,x^4 + 64\,x^6 )^7.
\end{eqnarray*}
For $r=1,3$ and 5, $P_{A_r,\,s}^{\Delta}(x)$ are of definite sign
in $-1< x<1$. They can never be orthogonal with each other for whichever
choice of the positive definite weight function.

%%%%%%%%%%%%%%%%%%%%%%%%%%%%
%  5.2                     %
%%%%%%%%%%%%%%%%%%%%%%%%%%%%
\subsection{$BC_r$ and $D_r$}

As shown in \cite{cs}, the equations (\ref{Wmax}) for $\Delta=BC_r$ read
\begin{equation}
   -2g_M\sum_{l=1\atop l\ne j}^r{\sin2\bar{q}_j\over
   {\cos2\bar{q}_j-\cos2\bar{q}_l}}+g_S{\cos\bar{q}_j\over{\sin \bar{q}_j}}
   +2g_L{\cos2\bar{q}_j\over{\sin2\bar{q}_j}}=0 \quad(j=1,\ldots,r).
\end{equation}
For non-vanishing $g_S$ and $g_L$, $\sin2\bar{q}_j=0$ cannot satisfy
the above equation. Thus by dividing by $\sin2\bar{q}_j$ we obtain for
$k_1\equiv g_S/g_M$, $k_2\equiv g_L/g_M$:
\begin{equation}
   \sum_{l=1\atop l\ne j}^r{1\over{\bar{x}_j-\bar{x}_l}}+
   {k_1+k_2\over{2(\bar{x}_j-1)}}
   +{k_2\over{2(\bar{x}_j+1)}}=0 \quad(j=1,\ldots,r),
\end{equation}
in which $\bar{x}_j\equiv\cos2\bar{q}_j$.
These are the equations satisfied by the zeros
$\{\bar{x}_j\,|\,j=1,\ldots,r\}$ of the Jacobi polynomial
$P_r^{(\alpha,\beta)}(x)$ \cite{szego} with
\begin{equation}
   \alpha=k_1+k_2-1, \quad \beta=k_2-1.
\end{equation}
The Rodrigues' formula for the Jacobi polynomial $P_n^{(\alpha,\beta)}(x)$
reads
\begin{eqnarray}
   P_n^{(\alpha,\beta)}(x)&\!\!=\!\!&
   {(-1)^n\over{2^nn!}}(1-x)^{-\alpha}(1+x)^{-\beta}
   \left({d\over{dx}}\right)^n\Bigl((1-x)^{n+\alpha}(1+x)^{n+\beta}\Bigr)\\
   &\!\!=\!\!&\frac{1}{2^nn!}
   \frac{\Gamma(2n+\alpha+\beta+1)}{\Gamma(n+\alpha+\beta+1)}x^n+\cdots.
   \nonumber
\end{eqnarray}

For $\Delta=D_r$, we have $g_S=g_L=0$, implying $\alpha=\beta=-1$. We choose
\[
   \bar{q}_1=0,\;\; \bar{q}_r=\pi/2\;\;(\,\Longleftrightarrow
   \cos2\bar{q}_1=1,\;\;\cos2\bar{q}_r=-1),
\]
then (\ref{Wmax}) read
\begin{equation}
   \sum_{l=2\atop l\ne j}^{r-1}
   {1\over{\bar{x}_j-\bar{x}_l}}+{1\over{\bar{x}_j-1}}
   +{1\over{\bar{x}_j+1}}=0 \quad (j=2,\ldots,r-1),
   \label{drroots}
\end{equation}
in which $\bar{x}_j\equiv\cos2\bar{q}_j$ ($j=2,\ldots,r-1$).
These are the equations satisfied by the zeros
$\{\bar{x}_j\,|\,j=2,\ldots,r-1\}$ of the Jacobi polynomial
$P_{r-2}^{(1,1)}(x)$ \cite{szego}. In fact, there is an identity
\begin{equation}
   4P_{r}^{(-1,-1)}(x)=(x^2-1)P_{r-2}^{(1,1)}(x),
   \label{jacobidr}
\end{equation}
which means that $\{1,\bar{x}_2,\ldots,\bar{x}_{r-1},-1\}$ are
the zeros of $P_r^{(-1,-1)}(x)$. This allows to treat $D_r$ as a limiting
case of $BC_r$.

The possible ${\cal R}$'s for $BC_r$ are $\Delta_S$, $\Delta_M$ and $\Delta_L$.
Since $\Delta_S=\{\pm{\bf e}_j\,|\,j=1,\ldots,r\}$ and
$\Delta_L=\{\pm2{\bf e}_j\,|\,j=1,\ldots,r\}$, we have trivial identities
among the polynomials
\begin{equation}
   P_{BC_r,\,s}^{\Delta_L}(k_1,\,k_2|x)=P_{BC_r,\,s2}^{\Delta_S}(k_1,\,k_2|x),
   \quad
   P_{BC_r,\,c}^{\Delta_L}(k_1,\,k_2|x)=P_{BC_r,\,c2}^{\Delta_S}(k_1,\,k_2|x).
\end{equation}
In other words, these relations prompted us to introduce the polynomials
of the forms $\prod_{\mu\in{\cal R}}\left(x-\sin(2\mu\cdot\bar{q})\right)$
and $\prod_{\mu\in{\cal R}}\left(x-\cos(2\mu\cdot\bar{q})\right)$.
For $BC_r$ Sutherland system we consider ${\cal R}=\Delta_S$ and $\Delta_M$
only.

%%%%%%%%%%%%%%%%%%%%%%%%%%%%
%  5.2.1                   %
%%%%%%%%%%%%%%%%%%%%%%%%%%%%
\subsubsection{${\cal R}=\Delta_S$ for $BC_r$}

Since $\Delta_S$ is {\em even\/} and that $\{\bar{x}_j=\cos2\bar{q}_j\,|\,
j=1,\ldots,r\}$ are the zeros of the Jacobi polynomial, it is natural to
consider the polynomial (\ref{sutpolycdefplus})
\begin{equation}
   P_{BC_r,\,c2}^{\Delta_{S+}}(k_1,k_2|x)
   =\prod_{j=1}^{r}\left(x-\cos2\bar{q}_j\right)
   =2^rr!\,{\Gamma(r+\alpha+\beta+1)\over{\Gamma(2r+\alpha+\beta+1)}}\,
   P_{r}^{(\alpha,\beta)}(x),
   \label{bcsjac}
\end{equation}
with $\alpha=k_1+k_2-1$ and $\beta=k_2-1$. They are orthogonal to each other:
\begin{equation}
   \int_{-1}^1 P_r^{(\alpha,\beta)}(x)
   P_s^{(\alpha,\beta)}(x)\,(1-x)^\alpha (1+x)^\beta\,dx\,
   \propto\delta_{r\,s}.
\end{equation}
As remarked in (\ref{sc2rel}), the polynomial
$P_{BC_r,\,c2}^{\Delta_{S+}}(k_1,k_2|x)$ %(\ref{bcsjac})
is equivalent with $P_{BC_r,\,s}^{\Delta_{S}}(k_1,k_2|x)$.
Needless to say that $2^nn!P_n^{(\alpha,\beta)}(x)$ is a
polynomial in the parameters $\alpha$ and $\beta$ with integer coefficients.
Thus $P_{BC_r,\,c2}^{\Delta_{S+}}(k_1,k_2|x)$ (\ref{bcsjac}) is a rational
function in $\alpha$ and $\beta$ with integer coefficients.

The other polynomial $P_{BC_r,\,s2}^{\Delta_{S}}(k_1,k_2|x)$
can be easily obtained by (\ref{s2c2rel}):
\begin{eqnarray}
   P_{BC_r,\,s2}^{\Delta_{S}}(k_1,k_2|x)&\!\!=\!\!&
   \prod_{j=1}^{r}\Bigl(x^2-\sin^2(2\bar{q}_j)\Bigr)
   \nonumber\\
   &\!\!=\!\!&(-1)^r\left(2^rr!\,
   {\Gamma(r+\alpha+\beta+1)\over\Gamma(2r+\alpha+\beta+1)}\right)^2\,
   P_{r}^{(\alpha,\beta)}(u)P_{r}^{(\beta,\alpha)}(u),
   \label{bcsjac2}
\end{eqnarray}
in which $u^2=1-x^2$. Remark that $P_{r}^{(\alpha,\beta)}(-x)=
(-1)^rP_{r}^{(\beta,\alpha)}(x)$.

%%%%%%%%%%%%%%%%%%%%%%%%%%%%
%  5.2.2                   %
%%%%%%%%%%%%%%%%%%%%%%%%%%%%
\subsubsection{${\cal R}={\bf V}$ for $D_r$}

This is a special ($k_1=k_2=0$ or $\alpha=\beta=-1$) case of the previous
example. As in the previous example, we introduce
\begin{eqnarray}
   &&P_r^{{\bf V}_+}(x)\equiv P_{D_r,\,c2}^{{\bf V}_+}(x)=
   \prod_{j=1}^{r}\left(x-\cos2\bar{q}_j\right)
   ={2^{r}r!(r-2)!\over{(2r-2)!}}P_{r}^{(-1,-1)}(x)\nonumber\\
   &&\qquad\quad=(x+1)(x-1)\prod_{j=2}^{r-1}\left(x-\bar{x}_j\right)
   ={2^{r-2}r!(r-2)!\over{(2r-2)!}}(x+1)(x-1)P_{r-2}^{(1,1)}(x).
   \label{jacobidens}
\end{eqnarray}
They are orthogonal to each other:
\begin{equation}
   \int_{-1}^1 P_r^{{\bf V}_+}(x)P_s^{{\bf V}_+}(x)\,(1-x)^{-1}(1+x)^{-1}\,dx\,
   \propto \int_{-1}^1 P_{r-2}^{(1,1)}(x)P_{s-2}^{(1,1)}(x)(1-x)(1+x)
   \,dx\,
   \propto\delta_{r\,s}.
\end{equation}
Corresponding to the  Dynkin diagram folding
$D_{r+1}\to B_r$ and (\ref{jacobidens}), we obtain
\begin{equation}
   (x-1)  P_{BC_r,\,c2}^{\Delta_{S+}}(2,0|x)=
   P_{D_{r+1},\,c2}^{{\bf V}_+}(x)=P_{BC_{r+1},\,c2}^{\Delta_{S+}}(0,0|x),
\end{equation}
which is the trigonometric counterpart of (\ref{drbrredV}).

The other polynomial $P_{D_r,\,s2}^{\bf V}(x)$ has a simple form
\begin{eqnarray}
   P_{D_r,\,s2}^{\bf V}(x)&\!\!=\!\!&
   \prod_{j=1}^{r}\Bigl(x^2-\sin^2(2\bar{q}_j)\Bigr)\nonumber\\
   &\!\!=\!\!&(-1)^r\left({2^{r-1}r!\,(r-2)!\over (2r-2)!}\right)^2\,
   x^4\left.\left(P_{r-2}^{(1,1)}(u)\right)^2\right|_{u^2\to 1-x^2},
   \label{drs2j}
\end{eqnarray}
which is of a definite sign in $-1< x< 1$. Thus they do not form any
orthogonal polynomials.

%%%%%%%%%%%%%%%%%%%%%%%%%%%%
%  5.2.3                   %
%%%%%%%%%%%%%%%%%%%%%%%%%%%%
\subsubsection{$A_{2r-1}\to C_r$ and the relationship between Chebyshev
and Jacobi polynomials}

As in the Calogero case, the Dynkin diagram folding $A_{2r-1}\to C_r$ implies
\begin{equation}
   P_{A_{2r-1},\,s}^{\bf V}(x)
   =(-2)^{-r}P_{BC_r,\,c2}^{\Delta_{S+}}(0,1/2|1-2x^2).
   \label{triga2r-1cr}
\end{equation}
Indeed there are relations between Chebyshev and Jacobi polynomials:
\begin{eqnarray}
   2^{1-2r}T_{2r}(x)&\!\!=\!\!&
   (-1)^r\frac{r!(r-1)!}{(2r-1)!}P_r^{(-1/2,-1/2)}(1-2x^2),
   \label{ChebJaco1}\\
   2^{-2r}T_{2r+1}(x)&\!\!=\!\!&
   (-1)^r\frac{(r!)^2}{(2r)!}xP_r^{(1/2,-1/2)}(1-2x^2),
   \label{ChebJaco2}
\end{eqnarray}
on top of the well-known relations (eq(4.1.7) of \cite{szego}):
\[
{1\cdot3\cdots(2r-1)\over{2\cdot4\cdots2r}}\,T_r(x)=
P_r^{(-1/2,-1/2)}(x).
\]
The former corresponds to (\ref{triga2r-1cr}) and the latter implies
\begin{equation}
   P_{A_{2r},\,s}^{\bf V}(x)
   =(-2)^{-r}xP_{BC_r,\,c2}^{\Delta_{S+}}(1,1/2|1-2x^2).
\end{equation}

%%%%%%%%%%%%%%%%%%%%%%%%%%%%
%  5.2.4                   %
%%%%%%%%%%%%%%%%%%%%%%%%%%%%
\subsubsection{${\cal R}={\bf S}$ and $\bar{\bf S}$ for $D_r$}

As in the Calogero systems, the symmetry of the $D_r$ Dynkin diagram
implies that $P_{D_{r},\,a}^{\bf S}(x)=P_{D_{r},\,a}^{\bar{\bf S}}(x)$,
$a=s,c,s2,c2$. Among them  $P_{D_{r},\,c}^{{\bf S},\,\bar{\bf S}}(x)$
do not always give rational polynomials. As remarked above (\ref{sc2rel}),
$P_{D_{r},\,s}^{{\bf S},\,\bar{\bf S}}(x)$ are equivalent to
$P_{D_{r},\,c2}^{{\bf S}_+,\,\bar{\bf S}_+}(x)$ for even rank $r$.
Thus we list for lower rank $r$ the polynomials
$P_{D_{r},\,c2}^{{\bf S}_+,\,\bar{\bf S}_+}(x)$ and
$P_{D_{r},\,s2}^{{\bf S},\,\bar{\bf S}}(x)$:
\begin{eqnarray}
   \hspace{-6mm}P_{D_{4},\,c2}^{{\bf V}_+,\,{\bf S}_+,\,\bar{\bf S}_+}(x)
   &\!\!=\!\!& (x^2-1)(x^2-1/5),\\
   \hspace{-6mm}P_{D_{4},\,s2}^{{\bf V},\,{\bf S},\,\bar{\bf S}}(x)&\!\!=\!\!&
   x^4(x^2-4/5)^2,\\
   \hspace{-6mm}P_{D_{5},\,c2}^{{\bf S},\,\bar{\bf S}}(x)&\!\!=\!\!&
   (x^2-1/2)^4(x^4-x^2-1/196)^2,
   \label{d5c2sp}\\
   \hspace{-6mm}P_{D_{5},\,s2}^{{\bf S},\,\bar{\bf S}}(x)&\!\!=\!\!&
   (x^2-1/2)^4(x^4-x^2-1/196)^2,\\
   \hspace{-6mm}P_{D_{6},\,c2}^{{\bf S}_+,\,\bar{\bf S}_+}(x)&\!\!=\!\!&
   3^{-4}7^{-3}x^4(21x^4-28x^2+8)^2(63x^4-72x^2+16),\\
   \hspace{-6mm}P_{D_{6},\,s2}^{{\bf S},\,\bar{\bf S}}(x)&\!\!=\!\!&
   3^{-8}7^{-6}(x^2-1)^4(21x^4-14x^2+1)^4(63x^4-54x^2+7)^2.
\end{eqnarray}
It is interesting to note that the formula (\ref{s2c2rel}) applies to
$D_5$ (conjugate) spinor representation ${\bf S}$ ($\bar{\bf S}$),
which is not {\em even\/}. This is because the set of values
$\{\mu\cdot\bar{q}\,|\,\mu\in{\bf S}\}$ is {\em even\/}. Moreover, the
function in (\ref{d5c2sp}) is invariant under $x^2\to 1-x^2$.

%%%%%%%%%%%%%%%%%%%%%%%%%%%%
%  5.2.5                   %
%%%%%%%%%%%%%%%%%%%%%%%%%%%%
\subsubsection{${\cal R}=\Delta_M$ for $BC_r$}
\label{DMBCr}

The set of middle roots is $\Delta_M=\{\pm({\bf e}_j-{\bf e}_l),
\pm({\bf e}_j+{\bf e}_l)\,|\,1\leq j<l\leq r\}$.
As in the Calogero systems in \ref{brrdell}, the polynomial
$P_{B_r,\, s}^{\Delta_M}(k|x)$ can be expressed neatly in terms of
the coefficients of the polynomial $P_{B_r,\,s}^{\Delta_S}(k|x)$.
Suppose we have two polynomials in $y$:
\begin{eqnarray}
    &&f=\prod_{i=1}^n(y-\sin^2x_i)=\sum_{i=0}^n(-1)^ia_i\,y^{n-i},\\
    &&g=\prod_{1\leq i<j\leq n}
    \Bigl(y-\sin^2(x_i-x_j)\Bigr)\Bigl(y-\sin^2(x_i+x_j)\Bigr).
\end{eqnarray}
Let us denote $b_i=\sin^2x_i$, then we obtain $g$ as a symmetric polynomial
in $b_i$:
\begin{equation}
   g=\prod_{1\leq i<j\leq n}\Bigl(y^2-2(b_i+b_j-2b_ib_j)y+(b_i-b_j)^2\Bigr),
\end{equation}
and $\{a_i\}$ are the basis of the  symmetric polynomials in $b_i$:
\begin{equation}
   a_i=\sum_{1\leq j_1<\cdots<j_i\leq n}b_{j_1}\cdots b_{j_i}.
\end{equation}
Thus $g$ can be expressed in terms of the coefficients $\{a_i\}$ of $f$
with integer coefficients. For example:
\begin{eqnarray}
   &&\hspace{-5mm}n=2\;:\;\;g=y^2-2(a_1-2a_2)y+a_1^2-4a_2,
   \label{tfgrel1}\\
   &&\hspace{-5mm}n=3\;:\nonumber\\
   &&\hspace{-5mm}\;\;
   g=y^6-4(a_1-a_2)y^5+2(3a_1^2-a_2-4a_1a_2-12a_3+8a_1a_3)y^4\nonumber\\
   &&\hspace{-5mm}\;\;\phantom{g=}
   -2(2a_1^3-a_1a_2-13a_3-2a_1^2a_2-4a_2^2-2a_1a_3+32a_2a_3-32a_3^2)y^3
   \nonumber\\
   &&\hspace{-5mm}\;\;\phantom{g=} +(a_1^4+2a_1^2a_2-7a_2^2-24a_1a_3
   -8a_1a_2^2-16a_1^2a_3+120a_2a_3+16a_1a_2a_3-144a_3^2)y^2\nonumber\\
   &&\hspace{-5mm}\;\;\phantom{g=}
   -2(a_1^3a_2-3a_1a_2^2-9a_1^2a_3+27a_2a_3-2a_2^3-2a_1^3a_3+18a_1a_2a_3
   -54a_3^2)y\nonumber\\
   &&\hspace{-5mm}\;\;\phantom{g=}
   +a_1^2a_2^2-4a_2^3-4a_1^3a_3+18a_1a_2a_3-27a_3^2.
   \label{tfgrel2}
\end{eqnarray}
If $f$ is of rational coefficients, so is $g$. 

Here are some explicit forms of $P_{BC_r,\,s}^{\Delta_M}(k_1,k_2|x)$ for 
lower rank $r$ (see also \cite{poly}):
\begin{equation}
   \hspace{-1mm}
   P_{BC_2,\,s}^{\Delta_M}(k_1,k_2|x)=
   \frac{4(1 + k_2)(1 + k_1 + k_2)}{(1 + k_1 + 2k_2)(2 + k_1 + 2k_2)^2}
   -\frac{4(1 + k_2)(1 + k_1 + k_2)}{(1 + k_1 + 2k_2)(2 + k_1 +2k_2)}y + y^2,
\end{equation}
\begin{eqnarray}
   P_{BC_3,\,s}^{\Delta_M}(k_1,k_2|x)&\!\!=\!\!&
   {108(1 + k_2)(2 + k_2)^2(1 + k_1 + k_2)(2 + k_1 + k_2)^2\over
        (2 + k_1 + 2k_2)^2(3 + k_1 + 2k_2)^3(4 + k_1 + 2k_2)^4}
   \nonumber\\
   && - {108(1 + k_2)(2 + k_2)^2(1 + k_1 + k_2)(2 + k_1 + k_2)^2
         (10 + 3k_1 + 6k_2)\over(2 + k_1 + 2k_2)^2(3 + k_1 + 2k_2)^3
         (4 + k_1 + 2k_2)^4} y\nonumber\\
   &&+ {9(2 + k_2)^2(2 + k_1 + k_2)^2\over(2 + k_1 + 2k_2)^2(3 + k_1 + 2k_2)^2
         (4 + k_1 + 2k_2)^4}
         \left(164 + 196k_1 + 41k_1^2\right.\nonumber\\
   && \qquad\qquad \left. + 392k_2 + 292k_1k_2 + 32k_1^2k_2 + 292k_2^2 +
          96k_1k_2^2 + 64k_2^3\right)y^2\nonumber\\
   &&  - {4(2 + k_2)(2 + k_1 + k_2)\over(2 + k_1 + 2k_2)^2
   (3 + k_1 + 2k_2)^2(4 + k_1 + 2k_2)^3}
         \left(792 + 1278k_1 + 639k_1^2\right.\nonumber\\
   &&\qquad\quad + 99k_1^3 + 2556k_2 + 3088k_1k_2 +
          1052k_1^2k_2 + 88k_1^3k_2 + 3088k_2^2 \nonumber\\
   &&\qquad\quad + 2562k_1k_2^2 +
          504k_1^2k_2^2 + 16k_1^3k_2^2 + 1708k_2^3 + 832k_1k_2^3 +
          64k_1^2k_2^3\nonumber\\
   &&\qquad\quad\left. + 416k_2^4 + 80k_1k_2^4 + 32k_2^5
      \right)y^3\nonumber\\
   && +
       {6(2 + k_2)(2 + k_1 + k_2)(26 + 17k_1 + 34k_2 + 8k_1k_2 + 8k_2^2)\over
        (2 + k_1 + 2k_2)(3 + k_1 + 2k_2)(4 + k_1 + 2k_2)^2}y^4\nonumber\\
   && -
       {12(2 + k_2)(2 + k_1 + k_2)\over(3 + k_1 + 2k_2)(4 + k_1 + 2k_2)}y^5
   + y^6.
\end{eqnarray}

%%%%%%%%%%%%%%%%%%%%%%%%%%%%
%  5.2.6                   %
%%%%%%%%%%%%%%%%%%%%%%%%%%%%
\subsubsection{${\cal R}=\Delta$ for $D_r$}

These are the $k_1=0$ and $k_2=0$ limit of the formulae given in the
previous subsection.
\begin{eqnarray}
   \hspace{-9mm}P_{D_{4},\,c2}^{{\Delta}_+}(x)&\!\!=\!\!&
   (1 + x)^3(-3/5 + x)(-1/5 +x^2)^4,\\
   \hspace{-9mm}P_{D_{4},\,s2}^{{\Delta}}(x)&\!\!=\!\!&
   x^6(-4/5 + x^2)^8(-16/25 + x^2),\\
   \hspace{-9mm}P_{D_{5},\,c2}^{{\Delta}_+}(x)&\!\!=\!\!&
   x^4(1 + x)^3(-1/7 + x)(-4/7 + x^2)^2(-3/7 + x^2)^4,\\
   \hspace{-9mm}P_{D_{5},\,s2}^{{\Delta}}(x)&\!\!=\!\!&
   (-1 + x^2)^4x^6(-4/7 + x^2)^8(-3/7 + x^2)^4(-48/49 + x^2),\\
   \hspace{-9mm}P_{D_{6},\,c2}^{{\Delta}_+}(x)&\!\!=\!\!&3^{-9}\,7^{-7}
   (1 + x)^4(-3 - 14x + 21x^2)(1 - 14x^2 + 21x^4)^4 (7 - 54x^2 + 63x^4)^2,\\
   \hspace{-9mm}P_{D_{6},\,s2}^{{\Delta}}(x)&\!\!=\!\!&3^{-18}\,7^{-14}
   x^8(8 - 28x^2 + 21x^4)^8(16 - 72x^2 + 63x^4)^4 (128 - 560x^2 + 441x^4).
\end{eqnarray}
It is trivial to verify that (\ref{s2c2rel}) are satisfied:
\begin{equation}
   P_{D_r,\,s2}^{{\Delta}}(x)
   =\left.\vT P_{D_r,\,c2}^{{\Delta}_+}(u)P_{D_r,\,c2}^{{\Delta}_+}(-u)
   \right|_{u^2\to1-x^2}.
\end{equation}
The Dynkin diagram folding $D_{r+1}\to B_r$ relates the functions
\begin{equation}
   P_{BC_{r,\,c2}}^{\Delta_{M+}}(2,0|x)
   \left(P_{BC_{r,\,c2}}^{\Delta_{S+}}(2,0|x)\right)^2=
   P_{D_{r+1,\,c2}}^{\Delta_+}(x)=P_{BC_{r+1,\,c2}}^{\Delta_{M+}}(0,0|x), 
   \label{tdrbrredDel}
\end{equation}
which is the  trigonometric counterpart of the identity (\ref{tdrbrredV}).

%%%%%%%%%%%%%%%%%%%%%%%%%%%%
%                          %
%%%%%%%%%%%%%%%%%%%%%%%%%%%%
\bigskip
Next we discuss the systems based on the exceptional root systems.
As in the Calogero systems, we have relied on numerical evaluation of
the equilibrium points.

%%%%%%%%%%%%%%%%%%%%%%%%%%%%
%  5.3                     %
%%%%%%%%%%%%%%%%%%%%%%%%%%%%
\subsection{$E_r$}

%%%%%%%%%%%%%%%%%%%%%%%%%%%%
%  5.3.1                   %
%%%%%%%%%%%%%%%%%%%%%%%%%%%%
\subsubsection{${\cal R}={\bf 27}$ and $\Delta$ for $E_6$}

We have evaluated two polynomials independently:
\begin{eqnarray}
   P_{E_6,\,c2}^{\bf 27}(x)&\!\!=\!\!&\prod_{\mu\in{\bf 27}}
   \Bigl(x-\cos(2\mu\cdot\bar{q})\Bigr)\nonumber\\
   &\!\!=\!\!& {(-1 + x)^3(1 + 2x)^6\over{2^{18}\,7^4\,11^6}}
   \left(-743 - 42651\,x + 708939\,x^2 - 1704045\,x^3 -
          1890504\,x^4\right.\nonumber\\[6pt]
   &&\left.\qquad\qquad  + 7043652\,x^5 + 1260336\,x^6
   - 9391536\,x^7 + 4174016\,x^9\right)^2,
   \label{e6t27c}
\end{eqnarray}
and
\begin{eqnarray}
   P_{E_6,\,s2}^{\bf 27}(x)&\!\!=\!\!&\prod_{\mu\in{\bf 27}}
   \Bigl(x-\sin(2\mu\cdot\bar{q})\Bigr)\nonumber\\
   &\!\!=\!\!& {x^3(-3 + 4x^2)^3\over{2^{18}\,7^4\,11^6}}
   \left(-221709312 + 39409774992\,x^2 -
         786312492840\,x^4\right.\nonumber\\
   &&\qquad  + 6804048466593\,x^6 - 32072860850184\,x^8 +
         89147361696624\,x^{10}\nonumber\\
   &&\qquad - 149154571577088\,x^{12} + 147001580732160\,x^{14} -
         78400843057152\,x^{16}\nonumber\\
   &&\qquad\left. + 17422409568256\,x^{18}\right).
\end{eqnarray}
Although the set of minimal weights $\bf 27$ is not {\em even\/}, that is
$-{\bf 27}=\overline{\bf 27}\neq{\bf 27}$, these two polynomials are related.
The ``formula (\ref{s2c2rel})" is valid,
\begin{equation}
   P_{E_6,\,s2}^{\bf 27}(x)
   =\left.\vTb \sqrt{P_{E_6,\,c2}^{\bf 27}(u)}
   \sqrt{P_{E_6,\,c2}^{\bf 27}(-u)}\right|_{u^2\to1-x^2}.
\end{equation}
This is the same situation encountered in the $D_5$ (conjugate) spinor
representations $\bf S$ ($\bar{\bf S}$) in (\ref{d5c2sp}).
This provides a strong support for the above results.

As for ${\cal R}=\Delta$, we have:
\begin{eqnarray}
   P_{E_6,\,c2}^{{\Delta}_+}(x)&\!\!=\!\!&\prod_{\rho\in{\Delta}_+}
   \Bigl(x-\cos(2\rho\cdot\bar{q})\Bigr)\nonumber\\
   &\!\!=\!\!& {(2x+1)^6\over{2^{24}\,7^7\,11^{11}}}
   \left( -235 - 627\,x + 231\,x^2 + 847\,x^3 \right) \nonumber\\
   &&\times
   \left( -743 - 42651\,x + 708939\,x^2 - 1704045\,x^3 -
        1890504\,x^4\right.\nonumber\\
   &&\left.\quad\ + 7043652\,x^5 + 1260336\,x^6 -
        9391536\,x^7 + 4174016\,x^9 \right)^3,
\end{eqnarray}
\begin{eqnarray}
   P_{E_6,\,s2}^{{\Delta}}(x)&=&\prod_{\rho\in{\Delta}_+}
   \Bigl(x^2-\sin(2\rho\cdot\bar{q})\Bigr)\nonumber\\
   &\!\!=\!\!&
   {\left( -3 + 4\,x^2 \right)^6\over{2^{48}\,7^{14}\,11^{22}}}\,
    \left( -48384 + 422928\,x^2 - 1036728\,x^4 +
      717409\,x^6 \right)\\
   &&\times
   \left( -221709312 + 39409774992\,x^2 -
        786312492840\,x^4+ 6804048466593\,x^6 \right.\nonumber\\
   &&\quad\ -
        32072860850184\,x^8 + 89147361696624\,x^{10}-
        149154571577088\,x^{12} \nonumber\\
   &&\left.\quad\ + 147001580732160\,x^{14} -
        78400843057152\,x^{16} + 17422409568256\,x^{18} \right)^3.\nonumber
\end{eqnarray}

%%%%%%%%%%%%%%%%%%%%%%%%%%%%
%  5.3.2                   %
%%%%%%%%%%%%%%%%%%%%%%%%%%%%
\subsubsection{${\cal R}={\bf 56}$ for $E_7$}

We have evaluated two polynomials independently:
\begin{eqnarray}
   P_{E_7,\,c2}^{{\bf 56}_+}(x)&\!\!=\!\!&\prod_{\mu\in{\bf 56}_+}
   \Bigl(x-\cos(2\mu\cdot\bar{q})\Bigr)\nonumber\\
   &\!\!=\!\!&{x^4\over{11^4\,13^5\,17^6}}\left(
   9332954265600 - 345319307827200\,x^2 +
    5422446428313600\,x^4\right.\nonumber\\
   &&\qquad\quad  - 47902580312348160\,x^6 +
    266584469614182720\,x^8\nonumber\\
   &&\qquad\quad  - 991356255189780480\,x^{10} +
    2543382104409514368\,x^{12}\nonumber\\
   &&\qquad\quad  -
    4564307435286703104\,x^{14} +
    5717674981551733200\,x^{16}\nonumber\\
   &&\qquad\quad  -
    4899020276961851040\,x^{18} +
    2736363552042360240\,x^{20}\nonumber\\
   &&\left.\qquad\quad  - 897719270582318184\,x^{22} +
    131214258464743597\,x^{24}\right),
\end{eqnarray}
and
\begin{eqnarray}
   P_{E_7,\,s2}^{{\bf 56}}(x)&=&\prod_{\mu\in{\bf 56}}
   \Bigl(x-\sin(2\mu\cdot\bar{q})\Bigr)\nonumber\\
   &\!\!=\!\!&{(-1 + x^2)^4\over{11^{8}\,13^{10}\,17^{12}}}\left(7824285157 -
   1019921980260\,x^2 +
          44927774191218\,x^4\right.\nonumber\\[6pt]
   &&\qquad\qquad  - 933762748148260\,x^6 + 10512912980210355\,x^8\nonumber\\
   &&\qquad\qquad   -
          70729109671077000\,x^{10} + 302444017343367900\,x^{12}\nonumber\\
   &&\qquad\qquad   -
          850322103495681960\,x^{14} + 1590230624766864795\,x^{16}\nonumber\\
   &&\qquad\qquad   -
          1957192223677842580\,x^{18} + 1521592634309937618\,x^{20}\nonumber\\
   &&\qquad\qquad  \left. -
          676851830994604980\,x^{22} + 131214258464743597\,x^{24}\right)^2.
\end{eqnarray}
These two polynomials satisfy (\ref{s2c2rel}),
\begin{equation}
   P_{E_7,\,s2}^{{\bf 56}}(x)=
   \left.\vT P_{E_7,\,c2}^{{\bf 56}_+}(u)P_{E_7,\,c2}^{{\bf 56}_+}(-u)
   \right|_{u^2\to1-x^2}.
\end{equation}

%%%%%%%%%%%%%%%%%%%%%%%%%%%%
%  5.3.3                   %
%%%%%%%%%%%%%%%%%%%%%%%%%%%%
\subsubsection{${\cal R}={\Delta}$ for $E_7$ and $E_8$}

The polynomials $P_{E_r,s(2)}^{\Delta}(x)$, $P_{E_r,c2}^{\Delta_+}(x)$,
$r=7,8$ are too long to be displayed here.
Their degrees are 63 and 126 for $E_7$ and 120 and 240 for $E_8$.
They are given in \cite{poly}. They all satisfy the consistency condition
(\ref{s2c2rel})
\begin{equation}
   P_{E_r,\,s2}^{{\Delta}}(x)=
   \left.\vT P_{E_r,\,c2}^{{\Delta}_+}(u)P_{E_r,\,c2}^{{\Delta}_+}(-u)
   \right|_{u^2\to1-x^2}\quad (r=6,7,8).
\end{equation}
at the level of each factor.

%%%%%%%%%%%%%%%%%%%%%%%%%%%%
%  5.4                     %
%%%%%%%%%%%%%%%%%%%%%%%%%%%%
\subsection{$F_4$}

We present the polynomials as a function of $k\equiv g_S/g_L$.
The polynomials $P_{F_4,\,c2}^{\Delta_{L+},\,\Delta_{S+}}(k|x)$ and
$P_{F_4,\,s2}^{\Delta_{L},\,\Delta_{S}}(k|x)$, satisfying the condition
(\ref{s2c2rel}), are too lengthy to be displayed here.
They are given in \cite{poly}.
Here we give $P_{F_4,\,s}^{\Delta_{L},\,\Delta_{S}}(k|x)$ which have shorter
forms. As before we use $y=x^2$.

%%%%%%%%%%%%%%%%%%%%%%%%%%%%
%  5.4.1                   %
%%%%%%%%%%%%%%%%%%%%%%%%%%%%
\subsubsection{${\cal R}=\Delta_L$ for $F_4$}

\begin{eqnarray}
   &&\hspace{-5mm}\;\;P_{4,\,s}^L(k|y)\equiv P_{F_4,\,s}^{\Delta_L}(k|x)
   =\prod_{\rho\in{\Delta_{L}}}\Bigl(x-\sin(\rho\cdot\bar{q})\Bigr)
   =\prod_{\rho\in{\Delta_{L+}}}\Bigl(y-\sin^2(\rho\cdot\bar{q})\Bigr)
   \nonumber\\
   &&\hspace{-8mm}
   ={2^{12}3^6(1 + k)^6(2 + k)^2(3 + k)^3(1 + 2k)\over
    {(3 + 2k)^3(4 + 3k)^4(5 + 3k)^5(6 + 5k)^6}}\nonumber\\
   &&\hspace{-5mm}\;\;-
     \ {2^{13}3^6(1 + k)^6(2 + k)^2(3 + k)^3(1 + 2k)(14 + 9k)\over{
        (3 + 2k)^3(4 + 3k)^4(5 + 3k)^5(6 + 5k)^6}}\,y\nonumber\\
   &&\hspace{-5mm}\;\;+\
       {2^{11}3^6(1 + k)^5(2 + k)^2(3 + k)^3(1 + 2k)(232 + 346k + 123k^2)
        \over{(3 + 2k)^2(4 + 3k)^4(5 + 3k)^5(6 + 5k)^6}}\,  y^2\nonumber\\
   &&\hspace{-5mm}\;\;- \
       {2^{11}3^4(1 + k)^5(2 + k)(3 + k)^3\over{(3 + 2k)^2(4 + 3k)^4
         (5 + 3k)^5(6 + 5k)^6}}\nonumber\\
   &&\hspace{-5mm}\;\;\quad\times
   \left(30432 + 133672k + 211560k^2 +
          155726k^3 + 54075k^4 + 7128k^5\right)\,y^3\nonumber\\
   &&\hspace{-5mm}\;\; +\, {2^{8}3^6(1 + k)^4(2 + k)(3 + k)^2
         \over{(3 + 2k)^2(4 + 3k)^3(5 + 3k)^4(6 + 5k)^6}}\nonumber\\
   &&\hspace{-5mm}\;\;\quad\times
   \left(19296 + 90360k
   + 159652k^2 + 137582k^3 + 61155k^4 + 13264k^5 +
          1088k^6\right)\,y^4\nonumber\\
   &&\hspace{-5mm}\;\; -
       {2^{9}3^4(1 + k)^4(2 + k)(3 + k)^2\over
        {(3 + 2k)^2(4 + 3k)^3(5 + 3k)^4(6 + 5k)^6}}
   \left(283824 + 1395972k + 2711556k^2\right.\nonumber\\
   &&\hspace{-5mm}\;\;\hspace{3cm}\left. +
          2704381k^3 + 1489217k^4 + 447066k^5 + 65952k^6 + 3456k^7\right)y^5
   \nonumber\\
   &&\hspace{-5mm}\;\;+
       {2^{7}3^4(1 + k)^3(3 + k)^2\over{(3 + 2k)^2(4 + 3k)^3
         (5 + 3k)^3(6 + 5k)^6}}\left(1046592 + 6283632k + 15907184k^2
   \right.\nonumber\\
   &&\hspace{-5mm}\;\;\hspace{3cm} +
          22205264k^3 + 18708264k^4 + 9754573k^5 + 3088726k^6
   \nonumber\\
   &&\hspace{-5mm}\;\; \hspace{3cm}\left.
   + 553392k^7+ 47232k^8 + 1152k^9\right)y^6\nonumber\\
   &&\hspace{-5mm}\;\; - {2^{8}3^4(1 + k)^3(3 + k)^2
         \over{(3 + 2k)^2(4 + 3k)^2(5 + 3k)^3
         (6 + 5k)^5}}\left(35736 + 163412k + 300546k^2\right.\nonumber\\
   &&\hspace{-5mm}\;\;\hspace{3cm}\left. + 286499k^3 + 151260k^4 + 43412k^5 +
          6048k^6 + 288k^7\right)y^7\nonumber\\
   &&\hspace{-5mm}\;\;+ {864(1 + k)^2(3 + k)\over
        {(3 + 2k)(4 + 3k)^2(5 + 3k)^2(6 + 5k)^4}}\nonumber\\
   &&\hspace{-5mm}\;\;\quad\times
   \left(33120 + 130392k + 199564k^2 +
          150034k^3 + 57649k^4 + 10632k^5 + 720k^6\right)y^8\nonumber\\
   &&\hspace{-5mm}\;\; -
       {1152(1 + k)^2(3 + k)\over{(3 + 2k)(4 + 3k)^2(5 + 3k)^2(6 + 5k)^3}}
   \nonumber\\
   &&\hspace{-5mm}\;\;\quad\times\left(3312 + 10668k + 12946k^2 + 7313k^3 +
          1899k^4 + 180k^5\right)y^9\nonumber\\
   &&\hspace{-5mm}\;\; + {144(1 + k)^2(3 + k)(116 + 133k + 30k^2)\over
        {(3 + 2k)(4 + 3k)(5 + 3k)(6 + 5k)^2}}
   y^{10}\nonumber\\
   &&\hspace{-5mm}\;\; -
       {72(1 + k)(3 + k)\over{(5 + 3k)(6 + 5k)}}y^{11} + y^{12}.
   \label{f4tl}
\end{eqnarray}

%%%%%%%%%%%%%%%%%%%%%%%%%%%%
%  5.4.2                   %
%%%%%%%%%%%%%%%%%%%%%%%%%%%%
\subsubsection{${\cal R}=\Delta_S$ for $F_4$}

\begin{eqnarray}
   &&\hspace{-5mm}\;\;P_{4,\,s}^S(k|y)\equiv P_{F_4,\,s}^{\Delta_S}(k|x)
   =\prod_{\rho\in{\Delta_{S}}}\Bigl(x-\sin(\rho\cdot\bar{q})\Bigr)
   =\prod_{\rho\in{\Delta_{S+}}}\Bigl(y-\sin^2(\rho\cdot\bar{q})\Bigr)
   \nonumber\\
   &&\hspace{-8mm}
   ={729k^3(1 + k)^6(2 + k)(3 + k)(1 + 2k)^2\over
        {(3 + 2k)^2(4 + 3k)^3(5 + 3k)^3(6 + 5k)^5}}\nonumber\\
   &&\hspace{-5mm}\;\; -
     {2916k^2(1 + k)^6(2 + k)(3 + k)(1 + 2k)^2(9 + 7k)\over
        {(3 + 2k)^2(4 + 3k)^3(5 + 3k)^3(6 + 5k)^5} }y\nonumber\\
   &&\hspace{-5mm}\;\; +
       {1458k(1 + k)^5(2 + k)(3 + k)(1 + 2k)^2(48 + 115k + 58k^2)\over
        {(3 + 2k)^2(4 + 3k)^2(5 + 3k)^3(6 + 5k)^5 }}\,y^2\nonumber\\
   &&\hspace{-5mm}\;\; -
       {324(1 + k)^5(3 + k)(1 + 2k)\over{(3 + 2k)^2(4 + 3k)^2
         (5 + 3k)^3(6 + 5k)^5}}\nonumber\\
   &&\hspace{-5mm}\;\;\quad\times
   \left(1152 + 11712k + 33125k^2 +
          38811k^3 + 20104k^4 + 3804k^5\right)\,y^3\nonumber\\
   &&\hspace{-5mm}\;\;
	   + {729(1 + k)^4(1 + 2k)
   (1536 + 8960k + 17519k^2 + 15049k^3 + 5788k^4 + 804k^5) \over
        {(3 + 2k)(4 + 3k)^2(5 + 3k)^3(6 + 5k)^4}}\, y^4\nonumber\\
   &&\hspace{-5mm}\;\;-
       {324(1 + k)^4(1 + 2k)\over{(3 + 2k)(4 + 3k)^2(5 + 3k)^3
         (6 + 5k)^4}}\nonumber\\
   &&\hspace{-5mm}\;\;\quad\times
   \left(26496 + 112704k + 177478k^2 + 130823k^3 +
          45354k^4 + 5913k^5\right)\,y^5\nonumber\\
   &&\hspace{-5mm}\;\;
	   + {162(1 + k)^3\over
       {(3 + 2k)(4 + 3k)^2(5 + 3k)^3(6 + 5k)^3}}\nonumber\\
   &&\hspace{-5mm}\;\;\quad\times
   \left(37824 + 208304k + 455436k^2 +
          505691k^3 + 300828k^4 + 90935k^5 + 10902k^6\right) \,y^6\nonumber\\
   &&\hspace{-5mm}\;\; -
      {324(1 + k)^3(9984 + 42832k + 70360k^2 + 55311k^3 + 20783k^4 +
          2978k^5)\over{(3 + 2k)(4 + 3k)^2(5 + 3k)^2(6 + 5k)^3}}\,y^7\nonumber\\
   &&\hspace{-5mm}\;\; +
      {54(1 + k)^2(4224 + 13765k + 16027k^2 + 7876k^3 + 1380k^4)\over
        {(3 + 2k)(4 + 3k)(5 + 3k)^2(6 + 5k)^2}}\,y^8 \nonumber\\
   &&\hspace{-5mm}\;\; -
       {72(1 + k)^2(2 + k)(345 + 628k + 276k^2)\over
        {(3 + 2k)(4 + 3k)(5 + 3k)(6 + 5k)^2}}\,y^9\nonumber\\
   &&\hspace{-5mm}\;\; +
      {36(1 + k)^2(52 + 29k)\over{(4 + 3k)(5 + 3k)(6 + 5k)}}\,y^{10} -
      {36(1 + k)\over{(6 + 5k)}}\,y^{11} + y^{12}
   \label{f4ts}
\end{eqnarray}
The folding $E_6\to F_4$ relates $E_6$ polynomials to $F_4$ polynomials at
the coupling ratio $k\equiv g_S/g_L=2$.
We have corresponding to (\ref{efredr272})
\begin{eqnarray}
   &&P_{F_4,\,s2}^{\Delta_S}(2|x)=P_{E_6,\,s2}^{\bf 27}(x)/x^3,\quad
   P_{F_4,\,c2}^{\Delta_S}(2|x)=P_{E_6,\,c2}^{\bf 27}(x)/(x-1)^3,
   \label{efredt272}\\
   &&P_{F_4,\,a}^{\Delta_L}(2|x)\left(P_{F_4,\,a}^{\Delta_S}(2|x)\right)^2=
   P_{E_6,\,a}^{\Delta}(x)\quad(a=s,s2),
   \label{efredt273}\\
   &&P_{F_4,\,c2}^{\Delta_{L+}}(2|x)
   \left(P_{F_4,\,c2}^{\Delta_{S+}}(2|x)\right)^2=
   P_{E_6,\,c2}^{\Delta_+}(x).
   \label{efredt274}
\end{eqnarray}
The self-duality of the $F_4$ Dynkin diagram relates $\Delta_L$ polynomials
to $\Delta_S$ ones. For example, we obtain:
\begin{eqnarray}
   &&\frac{847\,P_{4,\,s}^{L}(2|y)}{847y^3-1386y^2+594y-27}
   =\frac{64\,P_{4,\,s}^{S}(2|y)}{(4y-3)^3},
   \label{f4seld2}\\[5pt]
   &&\frac{717409\,P_{4,\,s2}^{L}(2|y)}
   {717409y^3-1036728y^2+422928y-48384}
   =\frac{64\,P_{4,\,s2}^{S}(2|y)}{(4y-3)^3},
   \label{f4seld1}\\[5pt]
   &&\frac{847\,P_{F_4,\,c2}^{\Delta_{L+}}(2|x)}{847x^3+231x^2-627x-235}
   =\frac{8\,P_{F_4,\,c2}^{\Delta_{S+}}(2|x)}{(2x+1)^3},
\end{eqnarray}
which are a factor of the parent polynomials $P_{E_6,\,s}^{\Delta}$, 
$P_{E_6,\,s2}^{\Delta}$ and $P_{E_6,\,c2}^{\Delta_+}$, respectively.

%%%%%%%%%%%%%%%%%%%%%%%%%%%%
%  5.5                     %
%%%%%%%%%%%%%%%%%%%%%%%%%%%%
\subsection{$G_2$}

Two types of polynomials
$\prod_{\rho\in{{\cal R}}_+}\left(x-\cos(2\rho\cdot\bar{q})\right)$ and
$\prod_{\rho\in{\cal R}}\left(x-\sin(2\rho\cdot\bar{q})\right)$ are evaluated.
For the latter we use $y=x^2$.

%%%%%%%%%%%%%%%%%%%%%%%%%%%%
%  5.5.1                   %
%%%%%%%%%%%%%%%%%%%%%%%%%%%%
\subsubsection{${\cal R}=\Delta_L$ for $G_2$}

\begin{eqnarray}
   P_{G_2,\,c2}^{\Delta_{L+}}(k|x)&\!\!=\!\!&\prod_{\rho\in{\Delta_{L+}}}
   \Bigl(x-\cos(2\rho\cdot\bar{q})\Bigr)\nonumber\\
   &\!\!=\!\!&\frac{27 - 81k - 99k^2 + 107k^3 + 80\,k^4 - 16\,k^5}
     {2(2 + k)^2\, (3 + 2\,k)^3} +
    \frac{3( 27 - 81\,k^2 - 40\,k^3 + 16\,k^4)}{2( 2 + k)(3 + 2\,k)^3}\,x
   \nonumber\\
   && + \frac{3(3 + 2\,k - 2\,k^2)} {( 2 + k)(3 + 2\,k)}\,x^2 + x^3,
\end{eqnarray}
\begin{eqnarray}
   P_{2,\,s2}^L(k|y)&\!\!\equiv\!\!&P_{G_2,\,s2}^{\Delta_{L}}(k|x)=
   \prod_{\rho\in{\Delta_{L}}}\Bigl(x-\sin(2\rho\cdot\bar{q})\Bigr)
   =\prod_{\rho\in{\Delta_{L+}}}\Bigl(y-\sin^2(2\rho\cdot\bar{q})\Bigr)
   \nonumber\\
   &\!\!=\!\!&-\frac{729\,{\left( 1 + k \right) }^2\,
       {\left( -3 + k + 8\,k^2 \right) }^2}{4\,
       {\left( 2 + k \right) }^4\,{\left( 3 + 2\,k \right) }^5}
   \nonumber\\
   &&
     + \frac{729\,{\left( 1 + k \right) }^2\,
       \left( 6 + 13\,k + 8\,k^2 \right) \,
       \left( 9 - 6\,k + 13\,k^2 + 8\,k^3 \right)}{4\,
       {\left( 2 + k \right) }^3\,{\left( 3 + 2\,k \right) }^6} \,y
   \nonumber\\
   &&
     - \frac{27\,\left( 1 + k \right) \,
       \left( 9 + 12\,k + 13\,k^2 + 8\,k^3 \right)}{
       {\left( 2 + k \right) }^2\,{\left( 3 + 2\,k \right) }^3} \,y^2
     + y^3,
\end{eqnarray}

%%%%%%%%%%%%%%%%%%%%%%%%%%%%
%  5.5.2                   %
%%%%%%%%%%%%%%%%%%%%%%%%%%%%
\subsubsection{${\cal R}=\Delta_S$ for $G_2$}

\begin{eqnarray}
   P_{G_2,\,c2}^{\Delta_{S+}}(k|x)&\!\!=\!\!&\prod_{\rho\in{\Delta_{S+}}}
   \Bigl(x-\cos(2\rho\cdot\bar{q})\Bigr)\nonumber\\
   &\!\!=\!\!&\frac{-9 - 21\,k - 13\,k^2 + k^3}
     {2(2 + k){( 3 + 2\,k)}^2} +
     \frac{3(-3 - 4\,k + k^2)}
     {2( 2 + k)( 3 + 2\,k)} \,x +
    \frac{3\,k}{3 + 2\,k}\,x^2 + x^3,
\end{eqnarray}
\begin{eqnarray}
   P_{2,\,s2}^S(k|y)&\!\!\equiv\!\!&P_{G_2,\,s2}^{\Delta_{S}}(k|x)
   =\prod_{\rho\in{\Delta_{S}}}\Bigl(x-\sin(2\rho\cdot\bar{q})\Bigr)
   =\prod_{\rho\in{\Delta_{S+}}}\Bigl(y-\sin^2(2\rho\cdot\bar{q})\Bigr)
   \nonumber\\
   &\!\!=\!\!&-\frac{27\,{\left( -3 + k \right) }^2\,k\,
       {\left( 1 + k \right) }^2}{4\,\left( 2 + k \right)%
       {\left( 3 + 2\,k \right) }^4} +
    \frac{27\,{\left( 1 + k \right) }^2\,
       \left( 9 + 12\,k + k^2 + 2\,k^3 \right)}{4\,
       {\left( 2 + k \right) }^2{\left( 3 + 2\,k \right) }^3}
   \,y\nonumber\\
   &&
     - \frac{9\,\left( 1 + k \right) \,
       \left( 3 + 2\,k + k^2 \right)}{\left( 2 + k
         \right){\left( 3 + 2\,k \right)}^2}\,y^2 + y^3.
\end{eqnarray}
They satisfy the formula (\ref{s2c2rel})
\begin{equation}
   P_{G_2,\,s2}^{{\Delta}_L}(x)=
   \left.\vT P_{G_2,\,c2}^{{\Delta}_{L+}}(u)P_{G_2,\,c2}^{{\Delta}_{L+}}(-u)
   \right|_{u^2\to1-x^2}\;,\quad
   P_{G_2,\,s2}^{{\Delta}_S}(x)=
   \left.\vT P_{G_2,\,c2}^{{\Delta}_{S+}}(u)P_{G_2,\,c2}^{{\Delta}_{S+}}(-u)
   \right|_{u^2\to1-x^2}\;.
\end{equation}
The Dynkin diagram folding $D_4\to G_2$ implies
\begin{eqnarray}
   &&\hspace{-5mm}
   P_{G_2,\,c2}^{\Delta_{S+}}(3|x)=P_{D_4,\,c2}^{{\cal R}_+}(x)/(x-1),\quad
   P_{G_2,\,s2}^{\Delta_S}(3|x)=P_{D_4,\,s2}^{{\cal R}}(x)/x^2
   \quad({\cal R}={\bf V},{\bf S},{\bf \bar{S}}),\\
   &&\hspace{-5mm}
   P_{G_2,\,a}^{\Delta_L}(3|x)\left(P_{G_2,\,a}^{\Delta_S}(3|x)\right)^3
   =P_{D_4,\,a}^{\Delta}(x)\quad(a=s,s2),
   \label{trid4g2}\\
   &&\hspace{-5mm}
   P_{G_2,\,c2}^{\Delta_{L+}}(3|x)
   \left(P_{G_2,\,c2}^{\Delta_{S+}}(3|x)\right)^3
   =P_{D_4,\,c2}^{\Delta_+}(x),
\end{eqnarray}
which correspond to (\ref{ratd4g2}).
The self-duality of the $G_2$ Dynkin diagram relates
$P_{G_2,\,s(2)}^{\Delta_L}(3|x)$ and
$P_{G_2,\,s(2)}^{\Delta_S}(3|x)$ (see \cite{poly} for
$P_{G_2,\,s}^{\Delta_{L,S}}(k|x)$):
\begin{eqnarray}
   &&\frac{5P_{2,\,s}^{L}(3|y)}{5y-1}=\frac{P_{2,\,s}^{S}(3|y)}{y-1},\quad
   \frac{25P_{2,\,s2}^{L}(3|y)}{25y-16}=\frac{P_{2,\,s2}^{S}(3|y)}{y},\\
   &&\frac{5P_{G_2,\,c2}^{\Delta_{L+}}(3|x)}{5x-3}
   =\frac{P_{G_2,\,c2}^{\Delta_{S+}}(3|x)}{x+1},
\end{eqnarray}
which are a factor of the parent polynomials, $P_{D_4,\,s}^{\Delta}$,
$P_{D_4,\,s2}^{\Delta}$ and $P_{D_4,\,c2}^{\Delta_+}$, respectively.

%%%%%%%%%%%%%%%%%%%%%%%%%%%%%%%%%%%%%%%%%%%%%%%%%%%%%%%%%%%%%%%
%                                                             %
%  6. Summary and Comments                                    %
%                                                             %
%%%%%%%%%%%%%%%%%%%%%%%%%%%%%%%%%%%%%%%%%%%%%%%%%%%%%%%%%%%%%%%
\section{Summary and Comments}
\label{comments}
\setcounter{equation}{0}

We have derived Coxeter (Weyl) invariant polynomials associated with
equilibrium points in Calogero and Sutherland systems based on all root
systems. For the classical root systems, the polynomials are well-known
classical orthogonal polynomials; Hermite, Laguerre, Chebyshev and Jacobi
of the degree equal to the rank $r$ of the root system ($r+1$ for the $A_r$
case), when the smallest set of weights ${\cal R}$ are chosen.
For the other choices of ${\cal R}$'s, the polynomials are related with
the corresponding classical polynomials but they no longer form an orthogonal
set. For the exceptional and non-crystallographic root systems,
these polynomials are new. Some polynomials are given in \cite{poly},
since they are too lengthy to be displayed in this paper.
These new polynomials have (much) higher degree than the rank $r$;
27 and 36 for $E_6$, 28 and 63 for $E_7$, 120 for $E_8$, 12 for $F_4$,
3 for $G_2$, $m$ for $I_2(m)$, 15 for $H_3$ and 60 for $H_4$.
Defined only for sporadic degrees, these new polynomials do not have
the orthogonality property, except for those corresponding to the dihedral
group $I_2(m)$ with the uniform coupling $g=g_e=g_o$.
In this case Chebyshev polynomials are obtained \cite{cs}.

All these new polynomials share one remarkable property with the classical
polynomials; their coefficients are rational functions of the ratio of
the coupling constants with all integer coefficients.
In most cases, they are monic polynomials with integer coefficients only.

In the rest of this section, we give a heuristic argument for ``deriving"
the classical orthogonal polynomials  with the proper weight function from
the pre-potential $W$ (\ref{Wform}) at equilibrium. We add one degree of
freedom, a new coordinate $q_{r+1}$ ($q_{r+2}$ for $A_r$), to the rank $r$
system at equilibrium:
\begin{equation}
   W(q_1,\ldots,q_r)\to
   \widetilde{W}(q_{r+1})=W(\bar{q}_1,\ldots,\bar{q}_r,q_{r+1}),
   \label{wtilde}
\end{equation}
and consider (rescaled) $q_{r+1}$ as the new variable. This is allowed
only for the classical root systems in which $r$ can be any positive integer.
Since ${\bf V}$ of $A_{r+2}$ has one more element $\mu_{r+2}$ than that 
of $A_r$, and $\Delta_S$ of $B_{r+1}$ ($BC_{r+1}$) has two more 
elements ${\bf e}_{r+1}$ and $-{\bf e}_{r+1}$ than that of $B_{r}$ ($BC_{r}$),
we multiply $\sqrt{dq_{r+2}}$ for $A_r$ case,
$(\sqrt{dq_{r+1}})^2=dq_{r+1}$ for $B_r$ ($BC_r$) case,
see (\ref{psiAr}), (\ref{psiBr}), (\ref{psiAr2}) and (\ref{psiBCr}).

%%%%%%%%%%%%%%%%%%%%%%%%%%%%
%  6.1                     %
%%%%%%%%%%%%%%%%%%%%%%%%%%%%
\subsection{Hermite}

The pre-potential for the $A_r$ Calogero system is
\[
   W=-\frac12\omega q^2+g\sum_{1\leq j<l\leq r+1}\log(q_j-q_l).
\]
After rescaling
\begin{equation}
q_{r+2}=\sqrt{g\over{\omega}}z,
\end{equation}
we obtain from (\ref{wtilde})
\begin{equation}
   \widetilde{W}(z)/g=
   -\frac12 z^2+\sum_{j=1}^{r+1}\log(z-\sqrt{\frac{\omega}{g}}\bar{q}_j)
   +(\mbox{$z$-indep.}).
\end{equation}
If we extract a function $\psi_{r+1}(z)$ from
\begin{eqnarray}
   e^{\widetilde{W}/g}\sqrt{dq_{r+2}}&\!\!=\!\!&(\mbox{$z$-indep.})\times
   e^{-z^2/2}\prod_{j=1}^{r+1}(z-\sqrt{\frac{\omega}{g}}\bar{q}_j)
   \times\sqrt{dz}\nonumber\\
   &\!\!=\!\!&(\mbox{$z$-indep.})\times e^{-z^2/2}H_{r+1}(z)\sqrt{dz}
   \nonumber\\
   &\!\!=\!\!&\psi_{r+1}(z)\sqrt{dz},
   \label{psiAr}
\end{eqnarray}
it satisfies the orthogonality relation
\[
   \int_{-\infty}^{\infty}dz\,\psi_n(z)\psi_m(z)\propto\delta_{n,m}.
\]

%%%%%%%%%%%%%%%%%%%%%%%%%%%%
%  6.2                     %
%%%%%%%%%%%%%%%%%%%%%%%%%%%%
\subsection{Laguerre}

The pre-potential for the $B_r$ Calogero system is
\[
   W=-\frac12\omega q^2+g_L\sum_{1\leq j<l\leq r}
   \log\Bigl((q_j-q_l)(q_j+q_l)\Bigr)+g_S\sum_{j=1}^r\log q_j.
\]
After rescaling
\begin{equation}
   q_{r+1}=\sqrt{{g_L\over{\omega}}z},
\end{equation}
we obtain from (\ref{wtilde})
\begin{equation}
   \widetilde{W}(z)/g_L= -\frac12z+\sum_{j=1}^r\log\Bigl(
   z-\Bigl(\sqrt{\frac{\omega}{g_L}}\bar{q}_j\Bigl)^2\Bigl)
   +\frac{k}{2}\log z+(\mbox{$z$-indep.}), \quad k\equiv g_S/g_L.
\end{equation}
If we extract a function $\psi_{r}(z)$ from
\begin{eqnarray}
   e^{\widetilde{W}/g_L}dq_{r+1}&\!\!=\!\!&(\mbox{$z$-indep.})\times
   z^{k/2}e^{-z/2}\prod_{j=1}^r\Bigl(
   z-\Bigl(\sqrt{\frac{\omega}{g_L}}\bar{q}_j\Bigr)^2\Bigr)
   \times z^{-1/2}dz\nonumber\\
   &\!\!=\!\!&(\mbox{$z$-indep.})\times
   z^{\alpha/2}e^{-z/2}L^{(\alpha)}_r(z)dz\nonumber\\
   &\!\!=\!\!&\psi_r(z)dz,\qquad \alpha\equiv k-1,
   \label{psiBr}
\end{eqnarray}
it satisfies the orthogonality relation
\[
   \int_0^{\infty}dz\,\psi_n(z)\psi_m(z)\propto\delta_{n,m}.
\]

%%%%%%%%%%%%%%%%%%%%%%%%%%%%
%  6.3                     %
%%%%%%%%%%%%%%%%%%%%%%%%%%%%
\subsection{Chebyshev}

This is slightly contrived.
The pre-potential for the $A_r$ Sutherland system is
\[
   W=g\sum_{1\leq j<l\leq r+1}\log\sin(q_j-q_l).
\]
By the choice of $\bar{q}$ (\ref{eqspaced}) and its property
$\bar{q}_j=-\bar{q}_{r+2-j}$, after defining
\begin{equation}
   \sin{q_{r+2}}=z,
\end{equation}
we obtain from (\ref{wtilde})
\begin{equation}
   \widetilde{W}(z)/g=\sum_{j=1}^{r+1}\log(z-\sin\bar{q}_j)+(\mbox{$z$-indep.}).
\end{equation}
If we extract a function $\psi_{r+1}(z)$ from
\begin{eqnarray}
   e^{\widetilde{W}/g}\sqrt{dq_{r+2}}&\!\!=\!\!&(\mbox{$z$-indep.})\times
   \prod_{j=1}^{r+1}(z-\sin\bar{q}_j)\times(1-z^2)^{-1/4}\sqrt{dz}\nonumber\\
   &\!\!=\!\!&(\mbox{$z$-indep.})\times
   (1-z^2)^{-1/4}T_{r+1}(z)\sqrt{dz}\nonumber\\
   &\!\!=\!\!&\psi_{r+1}(z)\sqrt{dz},
   \label{psiAr2}
\end{eqnarray}
it satisfies the orthogonality relation
\[
   \int_{-1}^1dz\,\psi_n(z)\psi_m(z)\propto\delta_{n,m}.
\]

%%%%%%%%%%%%%%%%%%%%%%%%%%%%
%  6.4                     %
%%%%%%%%%%%%%%%%%%%%%%%%%%%%
\subsection{Jacobi}

The pre-potential for the $BC_r$ Sutherland system is
\[
   W=g_M\sum_{1\leq j<l\leq r}\log\Bigl(\sin(q_j-q_l)\sin(q_j+q_l)\Bigr)
   +\sum_{j=1}^r(g_S\log\sin q_j+g_L\log\sin 2q_j).
\]
After defining $z$ by
\begin{equation}
   \cos2q_{r+1}=z,
\end{equation}
we obtain from (\ref{wtilde}) ($k_1\equiv g_S/g_M$, $k_2\equiv g_L/g_M$)
\begin{equation}
   \widetilde{W}(z)/g_M =
   \sum_{j=1}^r\log(z-\cos2\bar{q}_j)
   +\frac{k_1+k_2}{2}\log(1-z)+\frac{k_2}{2}\log(1+z)+(\mbox{$z$-indep.}).
\end{equation}
If we extract a function $\psi_{r}(z)$ from
\begin{eqnarray}
   e^{\widetilde{W}/g_M}dq_{r+1}&\!\!=\!\!&(\mbox{$z$-indep.})\times
   (1-z)^{(k_1+k_2)/2}(1+z)^{k_2/2}
   \prod_{j=1}^r(z-\cos 2\bar{q}_j)\times(1-z^2)^{-1/2}dz\nonumber\\
   &\!\!=\!\!&(\mbox{$z$-indep.})\times
   (1-z)^{\alpha/2}(1+z)^{\beta/2}P_r^{(\alpha,\beta)}(z)dz
   \nonumber\\
   &\!\!=\!\!&\psi_r(z)dz,\qquad\qquad
   \alpha\equiv k_1+k_2-1,\quad \beta\equiv k_2-1,
   \label{psiBCr}
\end{eqnarray}
it satisfies the orthogonality relation
\[
   \int_{-1}^1dz\,\psi_n(z)\psi_m(z)\propto\delta_{n,m}.
\]

%%%%%%%%%%%%%%%%%%%%%%%%%%%%%%%%%%%%%%%%%%%%%%%%%%%%%%%%%%%%%%%
%                                                             %
%  Acknowledgments                                            %
%                                                             %
%%%%%%%%%%%%%%%%%%%%%%%%%%%%%%%%%%%%%%%%%%%%%%%%%%%%%%%%%%%%%%%
\section*{Acknowledgements}
We thank  Toshiaki Shoji for useful discussion.
This work is supported in part by Grant-in-Aid for Scientific Research from
the Ministry of Education, Culture, Sports, Science and Technology,
No.12640261.

%%%%%%%%%%%%%%%%%%%%%%%%%%%%%%%%%%%%%%%%%%%%%%%%%%%%%%%%%%%%%%%
%                                                             %
%  References                                                 %
%                                                             %
%%%%%%%%%%%%%%%%%%%%%%%%%%%%%%%%%%%%%%%%%%%%%%%%%%%%%%%%%%%%%%%

\end{document}